\documentclass[aps,prx,twocolumn,floatfix]{revtex4-2}
\usepackage{graphicx}
\usepackage{multirow}
\usepackage{color}
\usepackage{amsmath}
\usepackage{amsfonts}
\usepackage{amssymb}
\usepackage{graphicx}
\graphicspath{{Figures/}}
\usepackage{epstopdf}
\usepackage{empheq}
\usepackage{float}
\usepackage{cases}
\usepackage{mathtools}
\usepackage{dcolumn}
\usepackage{bbold}
\usepackage{xfrac}
\usepackage{bm}
\usepackage{capt-of}

\usepackage[dvipsnames]{xcolor}
\usepackage[colorlinks]{hyperref}
\hypersetup{
colorlinks=True,
linkbordercolor=White,
linkcolor=BrickRed,
citecolor=Blue,	 
}

\DeclarePairedDelimiter\bra{\langle}{\rvert}
\DeclarePairedDelimiter\ket{\lvert}{\rangle}
\DeclarePairedDelimiter\braket{\langle}{\rangle}

\newcommand{\gt}{\mathsf{g}}

\newcommand\dexxeyydx{\frac{\partial(\varepsilon_{xx}-\varepsilon_{yy})}{\partial x}}
\newcommand\dexxeyydy{\frac{\partial(\varepsilon_{xx}-\varepsilon_{yy})}{\partial y}}
\newcommand\dexxeyydz{\frac{\partial(\varepsilon_{xx}-\varepsilon_{yy})}{\partial z}}
\newcommand\deyzdx{\frac{\partial\varepsilon_{yz}}{\partial x}}
\newcommand\deyzdy{\frac{\partial\varepsilon_{yz}}{\partial y}}
\newcommand\deyzdz{\frac{\partial\varepsilon_{yz}}{\partial z}}
\newcommand\dexzdx{\frac{\partial\varepsilon_{xz}}{\partial x}}
\newcommand\dexzdy{\frac{\partial\varepsilon_{xz}}{\partial y}}
\newcommand\dexzdz{\frac{\partial\varepsilon_{xz}}{\partial z}}
\newcommand\dexydx{\frac{\partial\varepsilon_{xy}}{\partial x}}
\newcommand\dexydy{\frac{\partial\varepsilon_{xy}}{\partial y}}
\newcommand\dexydz{\frac{\partial\varepsilon_{xy}}{\partial z}}

\newcommand\YMN[1]{#1}
\newcommand\YMNN[1]{#1}

\newcommand{\PRLsep}{\noindent\makebox[\linewidth]{\resizebox{0.5\linewidth}{1pt}{$\bullet$}}\bigskip}

\begin{document}

\title{Hole spin driving by strain-induced spin-orbit interactions}

\author{Jos\'e Carlos Abadillo-Uriel}
\author{Esteban A. Rodr\'iguez-Mena}
\author{Biel Martinez}
\author{Yann-Michel Niquet}
\email{yniquet@cea.fr}
\affiliation{Univ. Grenoble Alpes, CEA, IRIG-MEM-L\_Sim, Grenoble, France.}%

\date{\today}

\begin{abstract}
Hole spins in semiconductor quantum dots can be efficiently manipulated with radio-frequency electric fields owing to the strong spin-orbit interactions in the valence bands. Here we show that the motion of the dot in inhomogeneous strain fields gives rise to linear Rashba spin-orbit interactions (with spatially dependent spin-orbit lengths) and $\gt$-factor modulations that allow for fast Rabi oscillations. Such inhomogeneous strains build up spontaneously in \YMN{the devices} due to process and cool down stress. We discuss spin qubits in Ge/GeSi heterostructures as an illustration. We highlight that Rabi frequencies can be enhanced by one order of magnitude by shear strain gradients as small as $3\times 10^{-6}$\,nm$^{-1}$ within the dots. This underlines that spin in solids can be very sensitive to strains and opens the way for strain engineering in hole spin devices for quantum information and spintronics.
\end{abstract}

\maketitle

Hole spins in semiconductor quantum dots \cite{Burkard22} show versatile interactions with electric fields owing to the strong spin-orbit interaction (SOI) in the valence bands \cite{Winkler03,Kloeffel11,Kloeffel18}. This allows for fast electrical manipulation of hole spin qubits \cite{Maurand16,Crippa18,Watzinger18,Camenzind22,Froning21,wang2022ultrafast,Hendrickx20b} and for strong spin-photon interactions \cite{Kloeffel13,Bosco22,yu2022strong,michal2022tunable} suitable for long-range entanglement. The SOI, however, couples the spin to electrical and charge noise; yet recent works have shown how dephasing ``sweet spots'' can be engineered to limit decoherence \cite{Wang21,Bosco21,Piot22,michal2022tunable}. Ge/GeSi heterostructures have, in particular, made outstanding progress in the past two years \cite{Scappucci20,Hendrickx20b,Hendrickx20}, with the demonstration of a four qubits processor \cite{Hendrickx21} and of charge control in a sixteen dots array \cite{Borsoi22}.

The manipulation of hole spins by resonant AC electric fields involves a variety of physical manifestations of SOI. Rashba and Dresselhaus interactions couple the spin to the momentum of the hole, and give rise to an effective time-dependent magnetic field when the dot is shaken as a whole by the AC electric field \cite{Rashba03,Golovach06}. The modulations of the gyromagnetic $\gt$-factors of the hole resulting from the deformations of the moving dot may also drive spin rotations ($g$-tensor modulation resonance or $g$-TMR) \cite{Kato03,ares2013sige,Crippa18,Venitucci18}. The physics of SOI has been extensively investigated in Ge/GeSi heterostructures \cite{Marcellina17,Terrazos21,Bosco21b,Liu22,Salfi22,Bosco22b}; the role of the non-separability of the confinement potential and of the inhomogeneity of the AC electric field has in particular been highlighted \cite{martinez2022hole}. Yet the above mechanisms hardly seem sufficient to explain the large Rabi frequencies reported in some experiments \cite{Hendrickx20b,Hendrickx20,Hendrickx21}.

In this letter, we show that inhomogeneous strains give rise to specific linear Rashba and $g$-TMR mechanisms allowing for efficient electrical hole spin manipulation. We take Ge/GeSi heterostructures as an illustration, and demonstrate a tenfold increase in the Rabi frequencies for shear strain gradients as small as $3\times 10^{-6}$\,nm$^{-1}$, arising naturally from differential thermal contraction between materials \YMN{\cite{Thorbeck15}}. These mechanisms are likely ubiquitous in hole spin devices, but their fingerprints can easily be mingled with those of conventional (purely kinetic) Rashba SOI and $g$-TMR. This emphasizes how much spins in solids can be sensitive to strains \cite{Venitucci18,Morton18,Pla18,Liles20}. 

{\it Theory} -- We consider a hole moving in a potential $V(\mathbf{r})$ and a homogeneous magnetic field $\mathbf{B}$. The heavy-hole (HH) and light-hole (LH) Bloch functions can be mapped, respectively, onto the $J_z=\pm\tfrac{3}{2}$ and $J_z=\pm\tfrac{1}{2}$ components of a $J=\tfrac{3}{2}$ spin. The envelopes of these four Bloch functions fulfill a set of differential equations defined by the Luttinger-Kohn Hamiltonian \cite{Luttinger56,KP09}:
\begin{equation}
H=H_\mathrm{K}+H_\varepsilon+H_\mathrm{Z}+V(\mathbf{r})\mathbb{1}_4\,,
\label{eq:H4KP}
\end{equation}
where $H_\mathrm{K}$ is the kinetic energy, $H_\varepsilon$ describes the effects of strains, $H_\mathrm{Z}$ is the Zeeman Hamiltonian and $\mathbb{1}_4$ is the $4\times4$ identity matrix \footnote{We assume here holes with positive (electron-like) dispersion.}. $H_\mathrm{K}$ and $H_\varepsilon$ share the same generic form in the $J_z=\{+\tfrac{3}{2},+\tfrac{1}{2},-\tfrac{1}{2},-\tfrac{3}{2}\}$ basis set:
\begin{equation}
H_\mathrm{K/\varepsilon}=\begin{pmatrix}
P+Q & -S & R & 0 \\
-S^\dagger & P-Q & 0 & R \\
R^\dagger & 0 & P-Q & S \\
0 & R^\dagger & S^\dagger & P+Q
\end{pmatrix}
\label{eq:LK}\,,
\end{equation}
where, for $H_\mathrm{K}$, 
\begin{subequations}
\begin{align}
P_\mathrm{K}&=\frac{1}{2m_0}\gamma_1(p_x^2+p_y^2+p_z^2) \\
Q_\mathrm{K}&=\frac{1}{2m_0}\gamma_2(p_x^2+p_y^2-2p_z^2) \\
R_\mathrm{K}&=\frac{1}{2m_0}\sqrt{3}\left[-\gamma_2(p_x^2-p_y^2)+2i\gamma_3\{p_x,\,p_y\}\right] \\
S_\mathrm{K}&=\frac{1}{2m_0}2\sqrt{3}\gamma_3\{p_x-ip_y,\,p_z\}\,,
\end{align}
\end{subequations}
with $\{A,\,B\}=\tfrac{1}{2}(AB+BA)$, and, for $H_\varepsilon$,
\begin{subequations}
\begin{align}
P_\varepsilon&=-a_v(\varepsilon_{xx}+\varepsilon_{yy}+\varepsilon_{zz}) \\
Q_\varepsilon&=-\frac{1}{2}b_v(\varepsilon_{xx}+\varepsilon_{yy}-2\varepsilon_{zz}) \\
R_\varepsilon&=\frac{\sqrt{3}}{2}b_v(\varepsilon_{xx}-\varepsilon_{yy})-id_v\varepsilon_{xy} \\
S_\varepsilon&=-d_v(\varepsilon_{xz}-i\varepsilon_{yz})\,.
\end{align}
\end{subequations}
Here $\mathbf{p}$ is the momentum, $m_0$ is the free electron mass, and $\gamma_1$, $\gamma_2$, $\gamma_3$ are the Luttinger parameters that characterize the hole masses. The $\varepsilon_{\alpha\beta}$ are the strains; $a_v$ is the hydrostatic, $b_v$ the uniaxial and $d_v$ the shear deformation potential of the valence band. The form of Eq.~(\ref{eq:LK}), which couples different $J_z$'s through the $R$ and $S$ terms, embodies the action of SOI in the valence band. The Zeeman Hamiltonian $H_\mathrm{Z}=2\mu_B(\kappa\mathbf{B}\cdot\mathbf{J}+q\mathbf{B}\cdot\mathbf{J}^3)$ describes the action of the magnetic field on the Bloch functions, with $\mathbf{J}$ the spin $\tfrac{3}{2}$ operator, $\mathbf{J}^3\equiv(J_x^3,J_y^3,J_z^3)$, $\mu_B$ the Bohr magneton, and $\kappa$, $q$ the isotropic and cubic Zeeman parameters. The action of $\mathbf{B}$ on the envelopes of the hole is accounted for by the substitution $\mathbf{p}\to-i\hbar\boldsymbol{\nabla}+e\mathbf{A}$ in $H_\mathrm{K}$, with $\mathbf{A}=\frac{1}{2}\mathbf{B}\times\mathbf{r}$ the magnetic vector potential.

\begin{figure}
\includegraphics[width=.75\columnwidth]{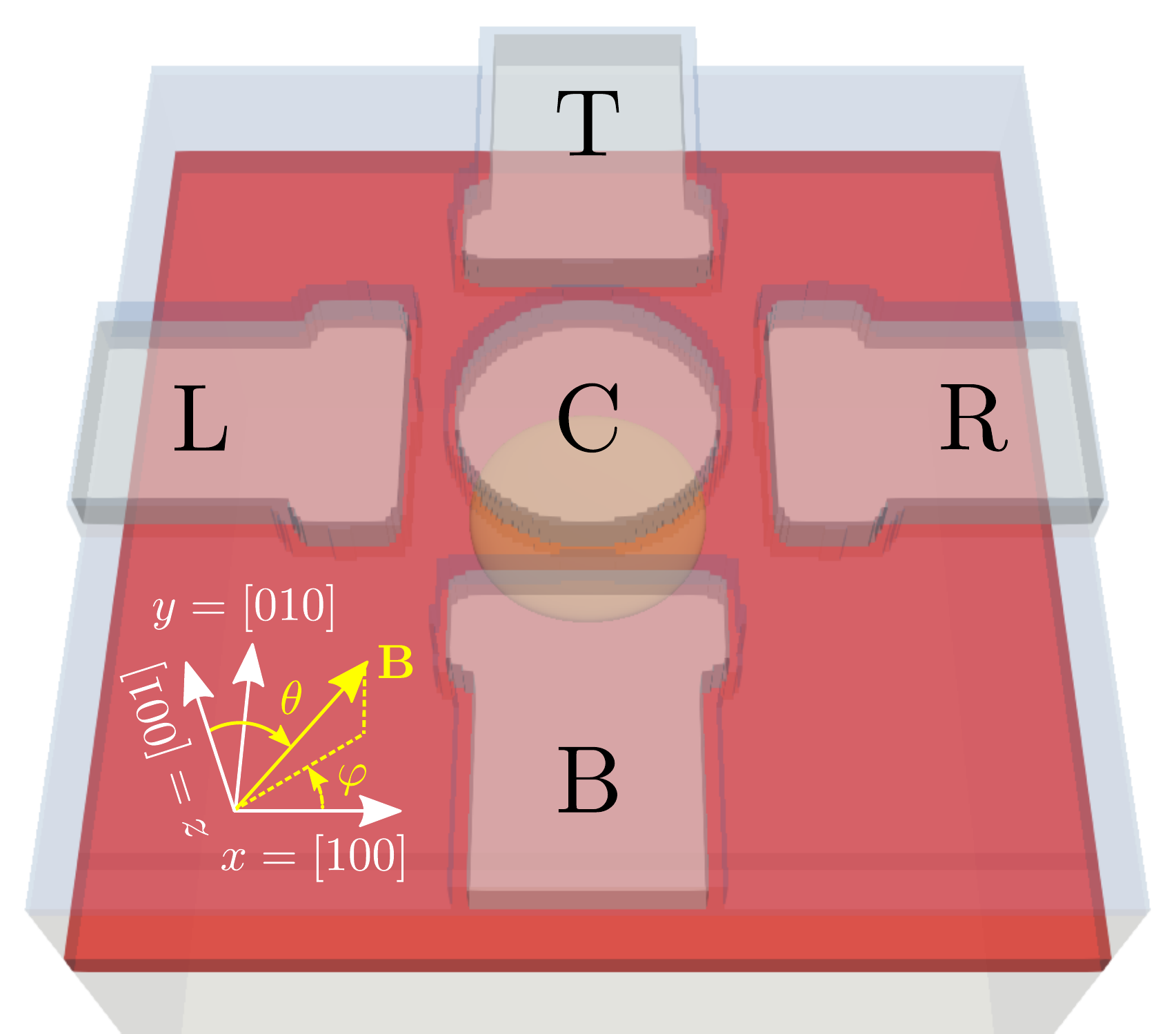}
\caption{The simulated device is made of 20\,nm thick Al gates (gray) on a Ge/Ge$_{0.8}$Si$_{0.2}$ heterostructure with a $L_\mathrm{W}=16$\,nm thick Ge well (red) and a 50\,nm thick upper GeSi barrier. The central C gate (diameter 100\,nm) is separated from the L/R/T/B side gates by 20\,nm. The gates are insulated from the substrate (and are surrounded on all facets) by 5\,nm of Al$_2$O$_3$ (blue). The yellow shape is the iso-density surface that encloses 90\% of the ground-state hole charge at bias $V_\mathrm{C}=-40$\,mV with side gates grounded.}
\label{fig:device}
\end{figure}

At $\mathbf{B}=\mathbf{0}$, the hole states are twofold degenerate owing to time-reversal symmetry. Each Kramers doublet splits at finite magnetic field and can be characterized by an effective Hamiltonian ${\cal H}=\tfrac{1}{2}\mu_B\boldsymbol{\sigma}\cdot\gt\mathbf{B}$ where $\boldsymbol{\sigma}$ is the vector of Pauli matrices and $\gt$ is the gyromagnetic $\gt$-matrix of the doublet \cite{Crippa18,Venitucci18}. We \YMNN{consider from now on a quantum dot strongly confined along $z=[001]$} (e.g., \YMNN{hosted} in a quantum well with thickness $L_\mathrm{W}$), although the following discussion can be extended to arbitrary structures. In the absence of HH/LH mixing [$R=S=0$ in Eq.~(\ref{eq:LK})], the ground-state is a pure $\{\ket{+\tfrac{3}{2}},\,\ket{-\tfrac{3}{2}}\}$ doublet split by $H_\mathrm{Z}$, with diagonal $\gt$-matrix ($\gt_{xx}=-\gt_{yy}=3q$, $\gt_{zz}=6\kappa+\tfrac{27}{2}q$). $R_\mathrm{K}$ and $S_\mathrm{K}$ actually admix LH components into the HH ground-state, owing, in particular, to lateral confinement in the $xy$ plane. The effects of this admixture on the $\gt$-matrix can be captured by a Schrieffer-Wolff \YMNN{(SW)} transformation \cite{Michal21}:
\begin{equation}
\delta{\cal H}_{hh^\prime}\approx\sum_{l}\frac{1}{E_h-E_l}\bra{h}H_\mathrm{c}\ket{l}\bra{l}H_\mathrm{c}^\prime\ket{h^\prime}\,,
\label{eq:SW}
\end{equation}
where $h$, $h^\prime$ run over the ground-state HH doublet with energy $E_{h^\prime}=E_h$, $l$ runs over LH states with energies $E_l$, and $H_\mathrm{c},\,H_\mathrm{c}^\prime\in\{R_\mathrm{K},S_\mathrm{K},H_\mathrm{Z}\}$. This yields \cite{Michal21,martinez2022hole}:
\begin{subequations}
\label{eq:gHH}
\begin{align}
\gt_{xx}&\approx+3q+\frac{6}{m_0\Delta_\mathrm{LH}}\left(\lambda\langle p_x^2\rangle-\lambda^\prime\langle p_y^2\rangle\right) \\
\gt_{yy}&\approx-3q-\frac{6}{m_0\Delta_\mathrm{LH}}\left(\lambda\langle p_y^2\rangle-\lambda^\prime\langle p_x^2\rangle\right) \\
\gt_{zz}&\approx6\kappa+\frac{27}{2}q-2\gamma_h\,,
\end{align}
\end{subequations}
where $\Delta_\mathrm{LH}=\frac{2\pi^2\hbar^2\gamma_2}{m_0L_\mathrm{W}^2}$ is the HH-LH band gap, $\lambda=\kappa\gamma_2-2\eta_h\gamma_3^2$, $\lambda^\prime=\kappa\gamma_2-2\eta_h\gamma_2\gamma_3$, $\gamma_h\approx 3.56$ and $\eta_h\approx 0.20$ in unstrained Ge films \cite{Ares13,Michal21,noteeta}. The expectations values of $p_x$ and $p_y$ are calculated for the ground-state HH envelope of the quantum dot. The $\propto\kappa\gamma_2$ contributions to $\gt_{xx}$ and $\gt_{yy}$ result from the interplay between $H_\mathrm{Z}$ and $R_\mathrm{K}$, while the $\propto\eta_h$ terms result from the action of the magnetic vector potential in $R_\mathrm{K}$ and the interplay with $S_\mathrm{K}$. We have assumed here $\langle p_\alpha p_\beta\rangle=0$ if $\alpha\ne\beta$ \cite{Michal21}.

The strain terms $R_\varepsilon$ and $S_\varepsilon$ also mix HH and LH states and give rise to $\gt$-matrix corrections. Neglecting orbital excitation energies with respect to the HH/LH band gap ($E_l-E_h\approx\Delta_\mathrm{LH}$), and using $\YMNN{\sum_l\braket{\mathbf{r}|l}\braket{l|\mathbf{r}^\prime}=\delta(\mathbf{r}-\mathbf{r}^\prime)}$, we get from the interplay between $H_\varepsilon$ and $H_\mathrm{Z}$:
\begin{subequations}
\label{eq:deltag1}
\begin{align}
\delta\gt_{xx}&=\delta\gt_{yy}=\frac{6b_v\kappa}{\Delta_\mathrm{LH}}\left(\langle\varepsilon_{yy}\rangle-\langle\varepsilon_{xx}\rangle\right)\\
\delta\gt_{zy}&=-\frac{4\sqrt{3}\kappa d_v}{\Delta_\mathrm{LH}}\langle\varepsilon_{yz}\rangle \\
\delta\gt_{zx}&=-\frac{4\sqrt{3}\kappa d_v}{\Delta_\mathrm{LH}}\langle\varepsilon_{xz}\rangle \\
\delta\gt_{xy}&=-\delta\gt_{yx}=\frac{4\sqrt{3}d_v\kappa}{\Delta_\mathrm{LH}}\langle\varepsilon_{xy}\rangle\,.
\end{align}
\end{subequations}
We have dropped the smaller $\propto q$ terms. Under biaxial strain $\varepsilon_{xx}=\varepsilon_{yy}=\varepsilon_\parallel$, $\varepsilon_{zz}=\varepsilon_\perp$, $\Delta_\mathrm{LH}\approx\frac{2\pi^2\hbar^2\gamma_2}{m_0L_\mathrm{W}^2}+2b_v(\varepsilon_\parallel-\varepsilon_\perp)$ but the above corrections are zero. Shear strains may bring non-zero off-diagonal elements in the $\gt$-matrix that rotate the principal magnetic axes as evidenced experimentally in Refs.~\onlinecite{Piot22,Liles20}.

\begin{figure*}
\includegraphics[width=\textwidth]{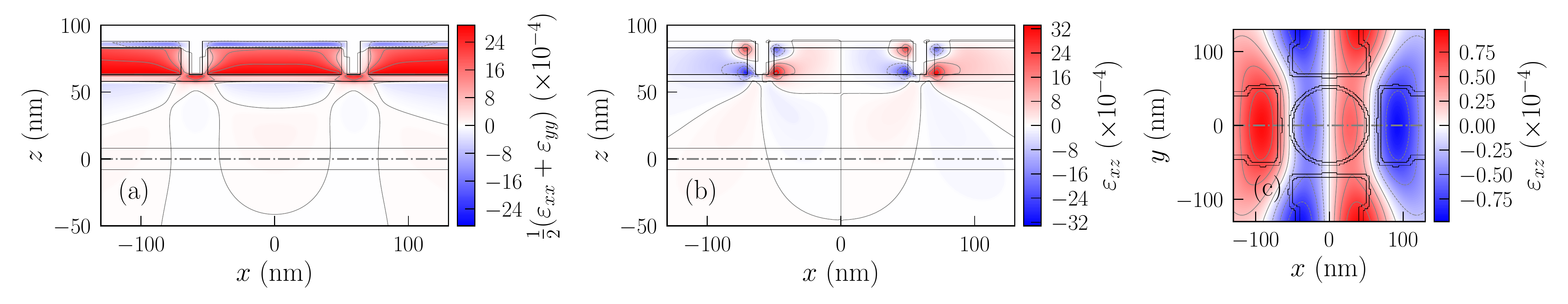}
\caption{Difference between inhomogeneous (with TC) and biaxial strains, in (a, b) the $xz$ plane at $y=0$, and (c) the $xy$ plane at $z=0$. These planes are identified by dashed-dotted gray lines in the panels. The black lines delineate the materials in (a, b) and the position of the gates (and Al$_2$O$_3$ around) at the surface of the heterostructure in (c). The strain $\varepsilon_{yz}$ in the Ge well is obtained from panel (c) by a $90^\circ$ rotation \cite{SM}. These maps are representative of the TC-induced strains in the device.}
\label{fig:strains}
\end{figure*}

Moreover, the interplay between $H_\mathrm{\varepsilon}$ and $H_\mathrm{K}$ gives rise to specific Rashba- and Dresselhaus-like SOIs. In particular, setting $H_\mathrm{c}=R_\mathrm{K}$, $H_\mathrm{c}^\prime=S_\mathrm{\varepsilon}$ (or vice-versa) yields
\begin{equation}
\delta{\cal H}_\mathrm{so}=-\frac{\hbar}{m_\parallel}\left[\frac{1}{\ell_\mathrm{so}}p_x-i\frac{\hbar}{2}\left(\frac{\partial}{\partial x}\frac{1}{\ell_\mathrm{so}}\right)\right]\sigma_y,   
\label{eq:rashba}
\end{equation}
with $m_\parallel$ the in-plane HH mass and $\ell_\mathrm{so}$ the spin-orbit length:
\begin{equation}
\frac{1}{\ell_\mathrm{so}}=\sqrt{3}\frac{m_\parallel d_v}{m_0\Delta_\mathrm{LH}}\left(\gamma_2\frac{\partial\varepsilon_{xz}}{\partial x}-\gamma_3\frac{\partial\varepsilon_{yz}}{\partial y}\right)\,.
\label{eq:lso}
\end{equation}
Note that $\ell_\mathrm{so}$ is generally dependent on position and signed (hence the $\propto\tfrac{\partial}{\partial x}(\ell_\mathrm{so}^{-1})$ correction for hermiticity) \cite{Dolcini18}. It is remarkable that inhomogeneous strains promote linear-in-momentum (instead of cubic) SOI even in symmetric dots. The complete set of strain-induced SOIs is given in the supplementary material \cite{SM}.

In general, $\gt$ is dependent on the gate voltages, which gives rise to Rabi oscillations when driving the dot with a resonant AC signal \cite{Kato03,Crippa18,Venitucci18}. The Rabi frequency reads:
\begin{equation}
f_\mathrm{R}=\frac{\mu_BBV_\mathrm{ac}}{2h\gt^*}\left|\gt\mathbf{b}\times\gt^\prime\mathbf{b}\right|\,,
\label{eq:fRg}
\end{equation}
where $\mathbf{b}$ is the unit vector along $\mathbf{B}$, $\gt^*=|\gt\mathbf{b}|$ is the effective $\gt$-factor of the dot, $V_\mathrm{ac}$ is the amplitude of the drive and $\gt^\prime=\tfrac{\partial}{\partial V}\gt$ is the derivative of $\gt$ with respect to the driven gate voltage. The latter collects different contributions \cite{Venitucci18}: Kinetic Rashba SOI \cite{Marcellina17,Terrazos21,Bosco21b}, also resulting from the interplay between $R_\mathrm{K}$ and $S_\mathrm{K}$ in Eq.~(\ref{eq:SW}), can give rise to non-zero off diagonal elements in $\gt^\prime$ when the dot is shaken as a whole \cite{Crippa18,Michal21}; the deformations of the dot in an anharmonic confinement potential and/or an inhomogeneous AC field directly modulate $\langle p_x^2\rangle$ and $\langle p_y^2\rangle$, hence $\gt_{xx}$, $\gt_{yy}$ and $\gt_{zz}$ (conventional $g$-TMR) \cite{Kato03,Crippa18,martinez2022hole}; the non-separability of the confinement in the $xy$ plane and along $z$ can result in rotations of the principal axes of the $\gt$-matrix and in non-zero $\gt_{zx}^\prime$ and $\gt_{zy}^\prime$ \cite{martinez2022hole}. Finally -- and this is the focus of this letter -- the motion and deformation of the dot in inhomogeneous strains can give rise to modulations of the $\delta\gt_{\alpha\beta}$'s [Eqs.~(\ref{eq:deltag1})] as well as to strain-induced Rashba SOI [Eq.~(\ref{eq:lso})].

{\it Application and discussion} -- As an illustration, \YMNN{we explore the contribution of these mechanisms to the Rabi oscillations of a hole spin qubit in a planar Ge/Ge$_{0.8}$Si$_{0.2}$ heterostructure \cite{Scappucci20,Hendrickx20b,Hendrickx20,Hendrickx21}}. We consider the device of Fig.~\ref{fig:device}, similar to Ref.~\onlinecite{martinez2022hole}. The quantum dot is shaped by the central C gate with the side L/R/T/B gates grounded. Practically, the C and side gates may be on different metalization levels \cite{Hendrickx21}; we keep, however, the structure as simple and symmetric as possible in order to best highlight the effects of strains. In the absence of the gate stack, the Ge well is biaxially strained by the Ge$_{0.8}$Si$_{0.2}$ buffer, with $\varepsilon_\parallel=-0.61\%$ and $\varepsilon_\perp=+0.45\%$. However, the Al gates and Al$_2$O$_3$ oxide imprint inhomogeneous strains resulting from fabrication and cool down. We assume here that the gate stack materials are nearly matched to the buffer at the temperature of their deposition ($T\approx 300$\,K for Al and $T\approx 550$\,K for Al$_2$O$_3$) and that inhomogeneous strains build up at $T\approx 0$\,K owing to the different thermal contraction (TC) coefficients (see \cite{SM} for details). This approach has been very successful in explaining the ESR lineshapes of Si:Bi substrates with Al resonators on top \cite{Pla18,Ranjan21}. The strains are calculated with a finite-element approach \cite{SM}.

The differences between inhomogeneous (with TC) and biaxial strains are plotted in Fig.~\ref{fig:strains} (see \cite{SM} for other strain components). The TC strains are mostly induced by the Al gates that contract much faster than the oxide and semiconductors. The effective lattice mismatch between the Al gates and Ge$_{0.8}$Si$_{0.2}$ buffer is indeed $\Delta a/a\approx-0.35\%$ at $T=0$\,K. The large $\varepsilon_\mathrm{p}=\tfrac{1}{2}(\varepsilon_{xx}+\varepsilon_{yy})\approx 0.27\%$ at the bottom interface of the C gate shows, however, that the contraction of Al is strongly hindered by the harder buffer and oxide. The strain modulations within the heterostructure are therefore small, with prominent shear components. They decrease with depth, reaching at most $\varepsilon_\mathrm{p}-\varepsilon_\parallel\approx 0.007\%$ in the Ge well. \YMN{We emphasize that the existence of such strains has been recently demonstrated experimentally in a similar layout \cite{Corley2023}.}

\begin{figure*}
\includegraphics[width=\textwidth]{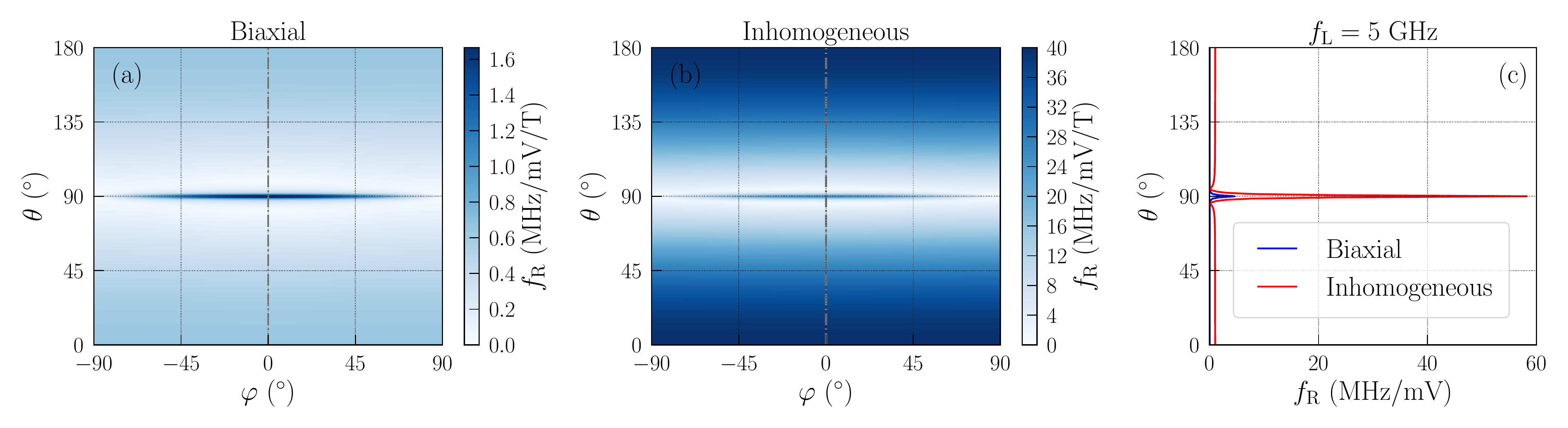}
\caption{(a, b) Maps of Rabi frequency as a function of the orientation of the magnetic field, for opposite drives $\delta V_\mathrm{L}=-\delta V_\mathrm{R}=\tfrac{1}{2}V_\mathrm{ac}\cos 2\pi f_\mathrm{L}t$ on the L and R gates ($V_\mathrm{C}=-40$\,mV). Map (a) is for homogeneous biaxial strains, and map (b) is with inhomogeneous TC strains. The Rabi frequency, proportional to $B$ and $V_\mathrm{ac}$, is normalized to $B=1$\,T and $V_\mathrm{ac}=1$\,mV. (c) Rabi frequency as a function of $\theta$ ($\varphi=0$) at constant Larmor frequency $f_\mathrm{L}=5$\,GHz, normalized to $V_\mathrm{ac}=1$\,mV.}
\label{fig:Rabi}
\end{figure*}

The electrical potential of the gates is computed with a finite-volumes method and the eigenstates of the dots with a finite-difference discretization of Eq.~(\ref{eq:H4KP}). The Rabi frequencies are then calculated from the numerical $\gt$-matrix and its derivative \cite{Venitucci18,SM}. \YMNN{This $\gt$-matrix formalism is non-perturbative in the HH/LH mixings and includes, therefore, all orders beyond Eqs.~(\ref{eq:deltag1}) and (\ref{eq:rashba})}. The maps of Rabi frequency as a function of the orientation of $\mathbf{B}$ are plotted in Fig.~\ref{fig:Rabi} for biaxial and inhomogeneous strains. The hole is driven by opposite AC modulations $\delta V_\mathrm{L}=-\delta V_\mathrm{R}=\tfrac{1}{2}V_\mathrm{ac}\cos2\pi f_\mathrm{L}t$ on the L and R gates, where $f_\mathrm{L}=\gt^*\mu_BB/h$ is the Larmor frequency (see \cite{SM} for drives with the L or C gate only). The maximal Rabi frequency (at constant $f_\mathrm{L}$) is enhanced by a factor $\approx 13$ by inhomogeneous CT strains. The anisotropy is nonetheless similar as in biaxial strains. \YMNN{Indeed, the first-order corrections $\delta\gt_{\alpha\beta}$ are all zero at $V_\mathrm{L}=V_\mathrm{R}=0$ given the symmetries of the device. The $\gt$-factors of the undriven dot are therefore almost the same in biaxial and in homogeneous strains}. Moreover, only $\gt_{xz}^\prime=\partial\gt_{xz}/\partial V_\mathrm{LR}$ and $\gt_{zx}^\prime=\partial\gt_{zx}/\partial V_\mathrm{LR}$ (with $V_\mathrm{LR}=V_\mathrm{L}-V_\mathrm{R}$) can be non-zero in both cases owing to the parity of the AC electric field \cite{Venitucci18,martinez2022hole,SM}. Therefore, for $\mathbf{B}$ in the $xz$ plane,
\begin{equation}
f_\mathrm{R}(\theta)=\frac{\mu_BBV_\mathrm{ac}}{2h\gt^*}\left|\gt_{xz}^\prime\gt_\perp\cos^2\theta-\gt_{zx}^\prime\gt_\parallel\sin^2\theta\right|\,,
\label{eq:frtheta}
\end{equation}
and for $\mathbf{B}$ in the $xy$ plane,
\begin{equation}
f_\mathrm{R}(\varphi)=\frac{\mu_BBV_\mathrm{ac}}{2h}\left|\gt_{zx}^\prime\cos\varphi\right|\,,
\end{equation}
where $\gt_\perp\equiv g_{zz}\approx 13.5$, $\gt_\parallel\equiv\gt_{xx}=-\gt_{yy}\approx 0.15$, and $\gt^*=\sqrt{\gt_\perp^2\cos^2\theta+\gt_\parallel^2\sin^2\theta}$ \cite{Wang22,martinez2022hole}. Thus $\gt_{xz}^\prime$ rules the out-of-plane, $\propto |b_z|$ background of Figs.~\ref{fig:Rabi}a,b, while $\gt_{zx}^\prime$ gives rise to the in-plane, $\propto |b_x|$ feature. The latter is particularly sharp (especially at constant $f_\mathrm{L}$) owing to the very large ratio between $\gt_\perp$ and $\gt_\parallel$. \YMNN{The mechanisms responsible for the Rabi oscillations in biaxial strains have been discussed in Ref.~\onlinecite{martinez2022hole}. The out-of-plane background ($\gt_{xz}^\prime=0.09$\,V$^{-1}$) stems from cubic Rashba SOI, while the in-plane feature ($\gt_{zx}^\prime=0.24$\,V$^{-1}$) is $g$-TMR resulting from the coupling between the motions along $x/y$ and $z$ in the non-separable confinement potential of the holes}. 

These mechanisms are superseded in inhomogeneous strains by the effects of the shear strains $\varepsilon_{xz}$ and $\varepsilon_{yz}$ (the other $\varepsilon_{\alpha\beta}$ making only minor contributions). The in-plane feature now picks the modulations of Eq.~(\ref{eq:deltag1}c) when the dot moves in the $\varepsilon_{xz}$ gradient. This is hence a $g$-TMR contribution, however leveraging the displacement $x_\mathrm{d}^\prime=\partial\langle x\rangle/\partial V_\mathrm{LR}$ of the dot rather than its deformations \cite{Kato03,Crippa18}. Using the calculated $x_\mathrm{d}^\prime=1.15$\,nm/mV and the biaxial HH/LH bandgap $\Delta_\mathrm{LH}\approx 71$\,meV, we estimate
$\delta\gt_{zx}^\prime\approx-4\sqrt{3}\kappa d_v x_\mathrm{d}^\prime\langle\tfrac{\partial}{\partial x}\varepsilon_{xz}\rangle/\Delta_\mathrm{LH}\approx 3.84$\,V$^{-1}$ \YMNN{from Eq.~(\ref{eq:deltag1}c). This is actually more than one decade larger than $\gt_{zx}^\prime=0.24$\,V$^{-1}$ in biaxial strains, and in fair agreement with the numerical (non-perturbative) $\gt_{zx}^\prime=3.25$\,V$^{-1}$, which shows that the SW transformation captures the main features of the strain-induced SOI}. The physics of the strain-induced Rashba SOI, Eq.~(\ref{eq:rashba}), is more intricate. If $\ell_\mathrm{so}$ is homogeneous (constant $\tfrac{\partial}{\partial x}\varepsilon_{xz}$ and $\tfrac{\partial}{\partial y}\varepsilon_{yz}$), $\delta{\cal H}_\mathrm{so}$ essentially couples the spin to the velocity $v_x=-(V_\mathrm{ac}x_\mathrm{d}^\prime) 2\pi f_\mathrm{L}\sin(2\pi f_\mathrm{L}t)$ of the driven hole, which results in a Rabi frequency $f_\mathrm{R}=V_\mathrm{ac}x_\mathrm{d}^\prime f_\mathrm{L}/\ell_\mathrm{so}$ when $\mathbf{B}\perp\mathbf{y}$ \cite{Golovach06}. In the $\gt$-matrix formalism, this translates into a small correction $-2\gt_\parallel x_\mathrm{d}^\prime/\ell_\mathrm{so}$ to $\gt_{zx}^\prime$, and into a sizable contribution $2\gt_\perp x_\mathrm{d}^\prime/\ell_\mathrm{so}$ to $\gt_{xz}^\prime$ \cite{SM}. However, when the spin-orbit lengths are inhomogeneous, the orbital motion of the hole in the magnetic vector potential becomes dependent on the dot position through the substitution $\mathbf{p}\to-i\hbar\boldsymbol{\nabla}+e\mathbf{A}$ in $\delta{\cal H}_\mathrm{so}$, which makes an even larger contribution to $\gt_{xz}^\prime$. From $\gt_{xz}^\prime=1.52$\,V$^{-1}$ without magnetic vector potential in $H_\mathrm{K}$, we estimate an effective $\ell_\mathrm{so}=2\gt_\perp x_\mathrm{d}^\prime/\gt_{xz}^\prime=32$\,$\mu$m, close to the expectation value of Eq.~(\ref{eq:lso}), $\ell_\mathrm{so}=40$\,$\mu$m; with the magnetic vector potential back on, $\gt_{xz}^\prime=5.70$\,V$^{-1}$ actually increases by a factor 4 \YMNN{(and is larger than the cubic Rashba contribution $\gt_{xz}^\prime=0.09$\,V$^{-1}$ by a factor 63). This large $\gt_{xz}^\prime$ can, however, hardly be harnessed efficiently because the magnetic field is much smaller along $z$ than in-plane at given $f_\mathrm{L}$ ($\gt_\perp\gg\gt_\parallel$). Rabi frequencies are practically larger for in-plane magnetic fields, and look more consistent with experimental data in inhomogeneous strains \cite{Hendrickx20,Hendrickx21} ($f_\mathrm{R}$ in the $50$\,MHz range indeed imply unreasonably large peak-to-peak modulations $2V_\mathrm{ac}\approx 20$\,mV in biaxial strains)}.

In the present device, the strain gradients are $\tfrac{\partial}{\partial x}\varepsilon_{xz}=\tfrac{\partial}{\partial y}\varepsilon_{yz}\approx 3\times10^{-6}$\,nm$^{-1}$ at the center of the dot. Residual shear strain gradients as small as $10^{-7}$\,nm$^{-1}$ would, therefore, still \YMNN{enhance significantly the} Rabi frequencies. \YMNN{We emphasize that the strains are primarily imposed by the same gates that shape the potential; they are therefore pervasive and commensurate with the dots, which strengthens their efficiency}. Also, $f_\mathrm{R}$ i s $\propto x_\mathrm{d}^\prime\propto r_\parallel^4$ for both strain-induced $g$-TMR and Rashba SOI, with $r_\parallel$ the radius of the dot. This is an unusually strong scaling for $g$-TMR contributions such as $\delta\gt_{zx}^\prime$ (Rashba SOI typically prevailing over purely kinetic $g$-TMR in long dots \cite{Michal21}). Strain-induced $g$-TMR shall, therefore, dominate over Rashba interactions whatever the size of the dot. Moreover, Fig.~\ref{fig:strains}b suggests that the Rabi oscillations speed up considerably if the Ge well is brought closer to the Al gates where shear strains are maximal. Calculations for a 25 nm thick Ge$_{0.8}$Si$_{0.2}$ barrier indeed show a $2.2\times$ enhancement of the Rabi frequencies \cite{SM}. \YMN{The prevalence of the above mechanisms can most easily be demonstrated experimentally by varying the nature or thickness of the metal gates, which has negligible impact on the electrostatics of a deeply buried well but modulates the strains in the heterostructure \cite{SM}}. \YMNN{Finally, we would like to outline the role of strain-induced SOI on the dephasing time $T_2^*$. Although stronger SOI is expected to decrease $T_2^*$, we find that inhomogeneously strained devices actually exhibit better quality factors $Q_2^*=2f_\mathrm{R}T_2^*$ over a wide range of magnetic field orientations thanks to the strong enhancement of the Rabi frequency $f_\mathrm{R}$. Moreover, biaxially and inhomogeneously strained devices display the same ``sweet spot'' $\mathbf{B}\parallel\mathbf{x}$ that maximizes $Q_2^*$ owing to symmetry and reciprocal sweetness relations between $f_\mathrm{R}$ and $T_2^*$ \cite{michal2022tunable}. Decoherence and relaxation are discussed in more details in the supplementary material \cite{SM}.}

To conclude, we have unveiled the specific linear Rashba SOI and $g$-TMR mechanisms arising from the motion of holes in inhomogeneous strain fields. In planar heterostructures, these mechanisms are essentially ruled by the gradients of shear strains $\varepsilon_{xz}$ and $\varepsilon_{yz}$. In Ge/GeSi spin qubits, they can make a prevalent contribution to the Rabi frequency even for the small shear strain gradients achieved by differential thermal contraction upon cool down. These mechanisms highlight the role of strains in spin-orbit physics and open the way for strain engineering in hole spin devices for quantum information \cite{Scappucci20}, hybrid semiconductor/superconductor and topological physics \cite{Maier14,Hendrickx18}, and spintronics \cite{Hirohata20,Lyandageller22}.

We thank R. Maurand for fruitful discussions and comments on the manuscript. This work was supported by the French National Research Agency (ANR) through the MAQSi project and the ``France 2030'' program (PEPR PRESQUILE-ANR-22-PETQ-0002).


\setcounter{section}{0}
\setcounter{equation}{0}
\setcounter{figure}{0}
\setcounter{table}{0}

\renewcommand\thefigure{S\arabic{figure}} 
\renewcommand\theequation{S\arabic{equation}}

\onecolumngrid

\vspace{1cm}
\PRLsep

\clearpage

\begin{center}
\textbf{\large Supplementary material for ``Hole spin driving by strain-induced spin-orbit interactions''}
\end{center}

\YMN{In this supplementary material, we give the material parameters (section \ref{sec:materials}) and the complete set of strains in the device of the main text (section \ref{sec:strains}). We next discuss Rabi oscillations driven by the L or C gate only (section \ref{sec:LCdrive}), as well as the impact of the thickness of the upper barrier (section \ref{sec:barrier}) and of the metal gates and oxide (section \ref{sec:metal}). We also give the full set of strain-induced spin-orbit interactions (section \ref{sec:deltags}), \YMNN{and discuss the coherence in Ge/GeSi heterostructures (section \ref{sec:coherence}). We finally address the calculation of numerical $\gt$-matrices as well as gauge invariance in the $\gt$-matrix formalism (section \ref{sec:gauge}), and derive the analytical expression of the $\gt$-matrix derivative in the presence of a Rashba spin-orbit interation (section \ref{sec:rashba})}}.

\section{Material parameters}
\label{sec:materials}

The material parameters are given in Table \ref{tab:params}. The lattice parameters of Si and Ge as a function of temperature are borrowed from Ref.~\onlinecite{Reeber96}. Those of Ge$_{0.8}$Si$_{0.2}$ are interpolated from the latter using Dismukes' law with a constant bowing $a[\mathrm{Ge}_{1-x}\mathrm{Si}_x]=(1-x)a[\mathrm{Ge}]+xa[\mathrm{Si}]-0.027\,\mathrm{\AA}\,x(1-x)$ \cite{Diskmukes64}. The elastic constants of Si and Ge at low temperature are from Refs.~\onlinecite{Hall67} and \onlinecite{Fine55}, and those of Ge$_{0.8}$Si$_{0.2}$ are linearly interpolated. Aluminium and Al$_2$O$_3$ are treated as isotropic elastic materials (due to their amorphous or granular nature); For Al, we compute the Lam\'e parameters $\lambda=61.37$\,GPa and $\mu=30.9$\,GPa from the bulk modulus $K=\lambda+2\mu/3=82$\,GPa and the shear modulus $G=\mu=30.9$\,GPa \cite{Vallin64}, from which we deduce isotropic $c_{11}=\lambda+2\mu=123.2$\,GPa, $c_{12}=\lambda=61.4$\,GPa and $c_{44}=\mu=30.9$\,GPa. For Al$_2$O$_3$, only room temperature data ara available. We compute the Lam\'e parameters $\lambda=63.3$ GPa and $\mu=68.5$ GPa from the Young modulus $E=\mu(3\lambda+2\mu)/(\lambda+\mu)=170$\,GPa and the Poisson ratio $\nu=\lambda/(2(\lambda+\mu))=0.24$ appropriate for atomic layer deposition (ALD) \cite{Tripp06,Ylivaara14}.

\begin{table}[h]
\centering
\begin{tabular}{l | r r r r | r r r r r r r r | r }
\toprule
 & \multicolumn{1}{r}{$a$ (\AA)} & \multicolumn{1}{r}{$c_{11}$ (GPa)} & \multicolumn{1}{r}{$c_{12}$ (GPa)} & \multicolumn{1}{r|}{$c_{44}$ (GPa)} & \multicolumn{1}{r}{$\gamma_1$} & \multicolumn{1}{r}{$\gamma_2$} & \multicolumn{1}{r}{$\gamma_3$} & \multicolumn{1}{r}{$a_v$ (eV)} & \multicolumn{1}{r}{$b_v$ (eV)} & \multicolumn{1}{r}{$d_v$ (eV)} & \multicolumn{1}{r}{$\kappa$} & \multicolumn{1}{r|}{$q$} & \multicolumn{1}{r}{$\varepsilon_\mathrm{r}$} \\
\hline
Si & 5.4298 & 167.5 & 64.9 & 80.2 & 4.285 & 0.339 & 1.446 & 2.10 & $-2.330$ & $-4.750$ & $-0.420$ & 0.01 & 11.7 \\
Ge & 5.6524 & 131.0 & 49.0 & 68.8 & 13.380 & 4.240 & 5.690 & 2.00 & $-2.160$ & $-6.060$ & 3.410 & 0.06 & 16.2  \\
Ge$_{0.8}$Si$_{0.2}$ & 5.6035 & 138.3 & 52.2 & 71.1 & 11.561 & 3.460 & 4.841 & 2.02 & $-2.194$ & $-5.798$ & 2.644 & 0.05 & 15.3 \\
Al$_2$O$_3$ & 5.6129 & 200.3 & 63.3 & 68.5 & \multicolumn{1}{r}{-} & \multicolumn{1}{r}{-} & \multicolumn{1}{r}{-} & \multicolumn{1}{r}{-} & \multicolumn{1}{r}{-} & \multicolumn{1}{r}{-} & \multicolumn{1}{r}{-} & \multicolumn{1}{r|}{-} & 8.0 \\
Al & 5.5985 & 123.2 & 61.4 & 30.9 & \multicolumn{1}{r}{-} & \multicolumn{1}{r}{-} & \multicolumn{1}{r}{-} & \multicolumn{1}{r}{-} & \multicolumn{1}{r}{-} & \multicolumn{1}{r}{-} & \multicolumn{1}{r}{-} & \multicolumn{1}{r|}{-} & \multicolumn{1}{r}{-} \\
\hline
\botrule
\end{tabular}
\caption{(Effective) lattice parameter $a$ and elastic constants $c_{11}$, $c_{12}$ and $c_{44}$ of the different materials at $T=0$\,K; Luttinger parameters $\gamma_1$, $\gamma_2$, $\gamma_3$, valence band deformation potentials $a_v$, $b_v$ and $d_v$, and Zeeman parameters $\kappa$ and $q$; \YMN{dielectric constant $\varepsilon_\mathrm{r}$}.}
\label{tab:params}
\end{table}

The lattice parameter of the Ge$_{0.8}$Si$_{0.2}$ buffer is $a=5.6035$\,\AA\ at $T=0$\,K, $a=5.6083$\,\AA\ at $T=300$\,K (typical deposition temperature of Al) and $a=5.6163$\,\AA\ at $T=550$\,K (typical ALD temperature for Al$_2$O$_3$). The corresponding thermal contraction (TC) coefficients at $T=0$\,K are therefore $\Delta a/a=[a(0\,\mathrm{K})-a(T)]/a(T)=-0.855\times 10^{-3}$ from $T=300$\,K, and $\Delta a/a=-2.279\times 10^{-3}$ from $T=550$\,K. We assume that there is however a residual strain $\varepsilon_{\parallel}=0.26\%$ in the buffer \cite{Sammak19}, roughly independent on temperature \footnote{This residual strain, measured at room temperature, likely results from the difference of thermal expansion coefficients between the buffer and the Si substrate down from the growth temperature $T\approx 1000$\,K. We may thus alternatively assume that the Si substrate still rules the thermal contraction of the buffer down to $T=0$\,K, which yields $\Delta a/a=-0.233\times 10^{-3}$ from $T=300$\,K, and an even larger $\varepsilon_{\mathrm{Al}}$.}, and that Al is deposited unstrained on this buffer at $T=300$\,K. Given the TC coefficient of Al, $\Delta l/l=[l(0\,\mathrm{K})-l(300\,\mathrm{K})]/l(300\,\mathrm{K})=-4.333\times 10^{-3}$ \cite{Nix41}, the effective lattice parameter at $T=0$\,K, used as input for the finite-elements calculation, is therefore $a[\mathrm{Al}]=5.6083(1+0.26\%)(1+\Delta l/l)=5.5985$\,\AA. The net lattice mismatch with the residually strained buffer is hence $\varepsilon_\mathrm{Al}=-0.35\%\approx\Delta l/l-\Delta a/a$. As for Al$_2$O$_3$, we assume likewise ALD at $T=550$\,K with a residual in-plane stress $\sigma=200$\,MPa \cite{Ylivaara14}. From the linear thermal expansion coefficient of Al$_2$O$_3$, $\alpha=4.2\times 10^{-6}$/K \cite{Miller10}, we estimate an effective lattice parameter $a[\mathrm{Al}_2\mathrm{0}_3]=5.6163(1+0.26\%)(1-\sigma(1-\nu)/E-\alpha\times(550\,\mathrm{K}))=5.6129$\,\AA\ at $T=0$\,K, and a net lattice mismatch with the buffer $\varepsilon_{\mathrm{Al}_2\mathrm{O}_3}=-0.09\%$. The use of a constant thermal expansion coefficient for Al$_2$O$_3$ may be questioned; however the data for this material are pretty scattered at room temperature (due to is amorphous nature), and not available at low temperature. The thin aluminium oxide has, nonetheless, little impact on the strain distributions; the TC stress is, indeed, dominated by the strong contraction $\Delta l/l$ of Aluminium with respect to the oxide and semiconductors. 

The Luttinger and Zeeman parameters of Si and Ge are from Ref.~\onlinecite{Winkler03}, and the valence band deformation potentials from Ref.~\onlinecite{Fischetti96}. The electronic parameters of Ge$_{0.8}$Si$_{0.2}$ are linearly interpolated. The band offset between unstrained Ge$_{0.8}$Si$_{0.2}$ and Ge is $\Delta_\mathrm{VBO}=0.138$\,eV.

\begin{figure}
\centering
\includegraphics[width=.45\columnwidth]{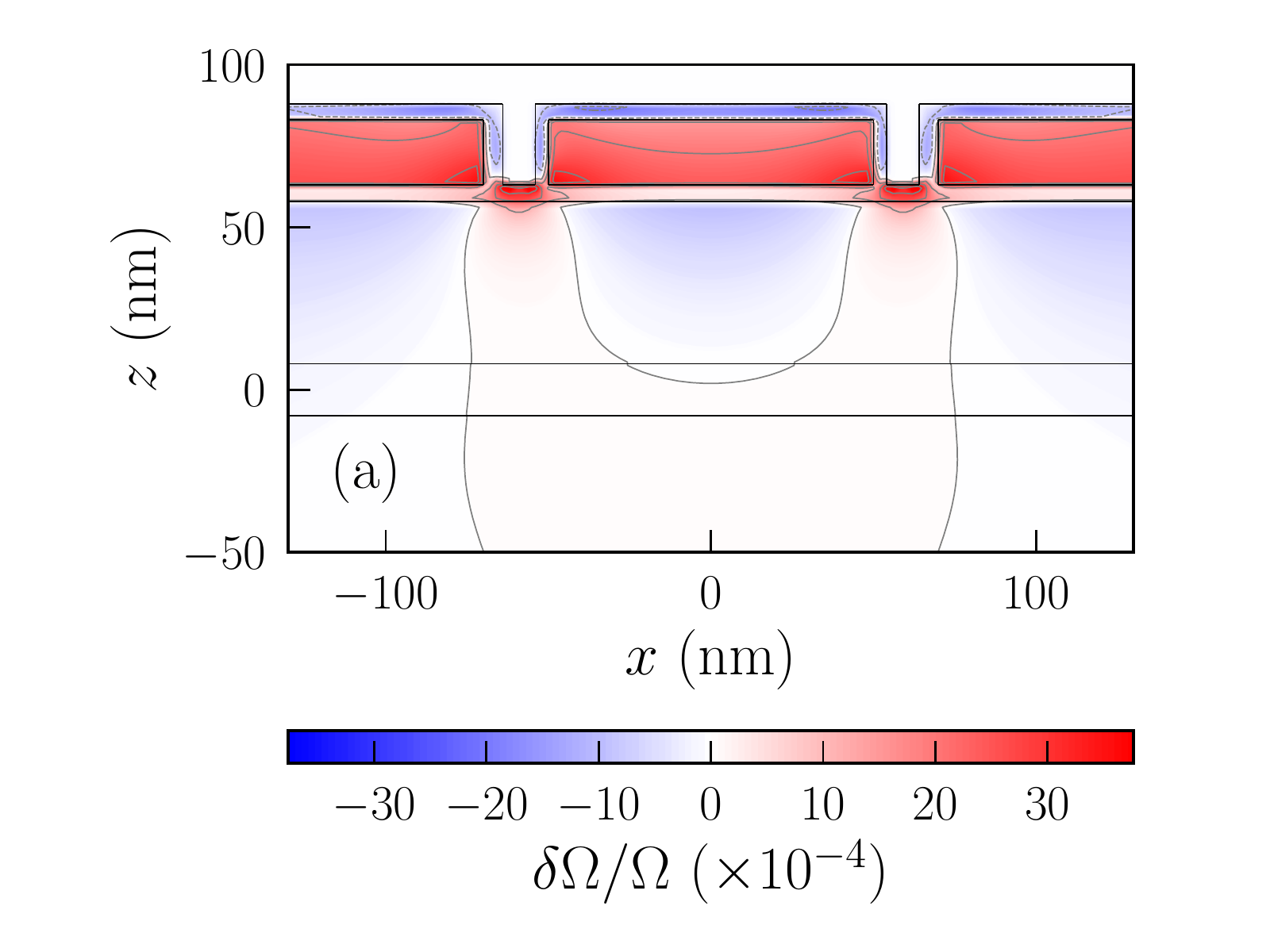}
\includegraphics[width=.45\columnwidth]{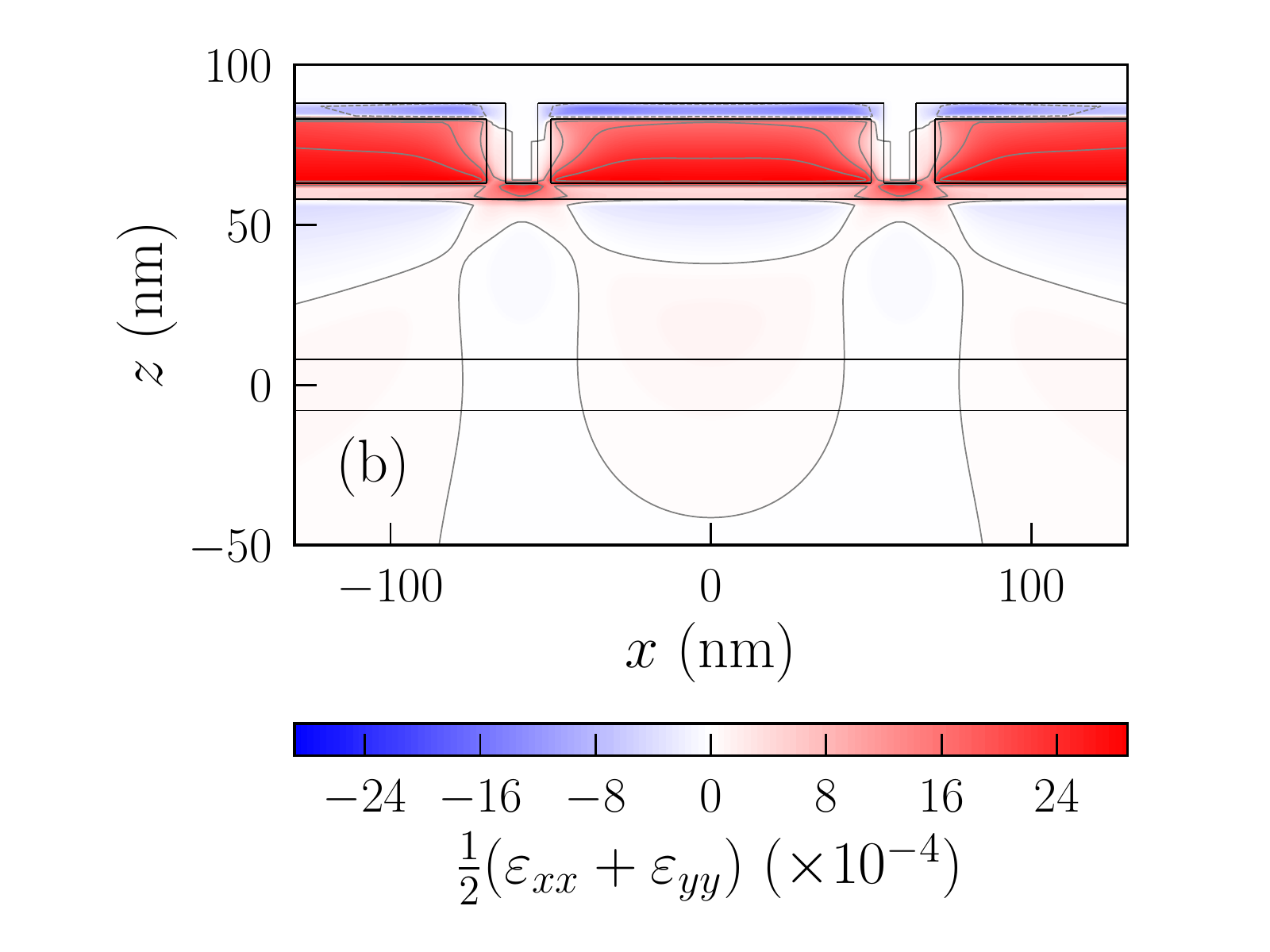} \\
\includegraphics[width=.45\columnwidth]{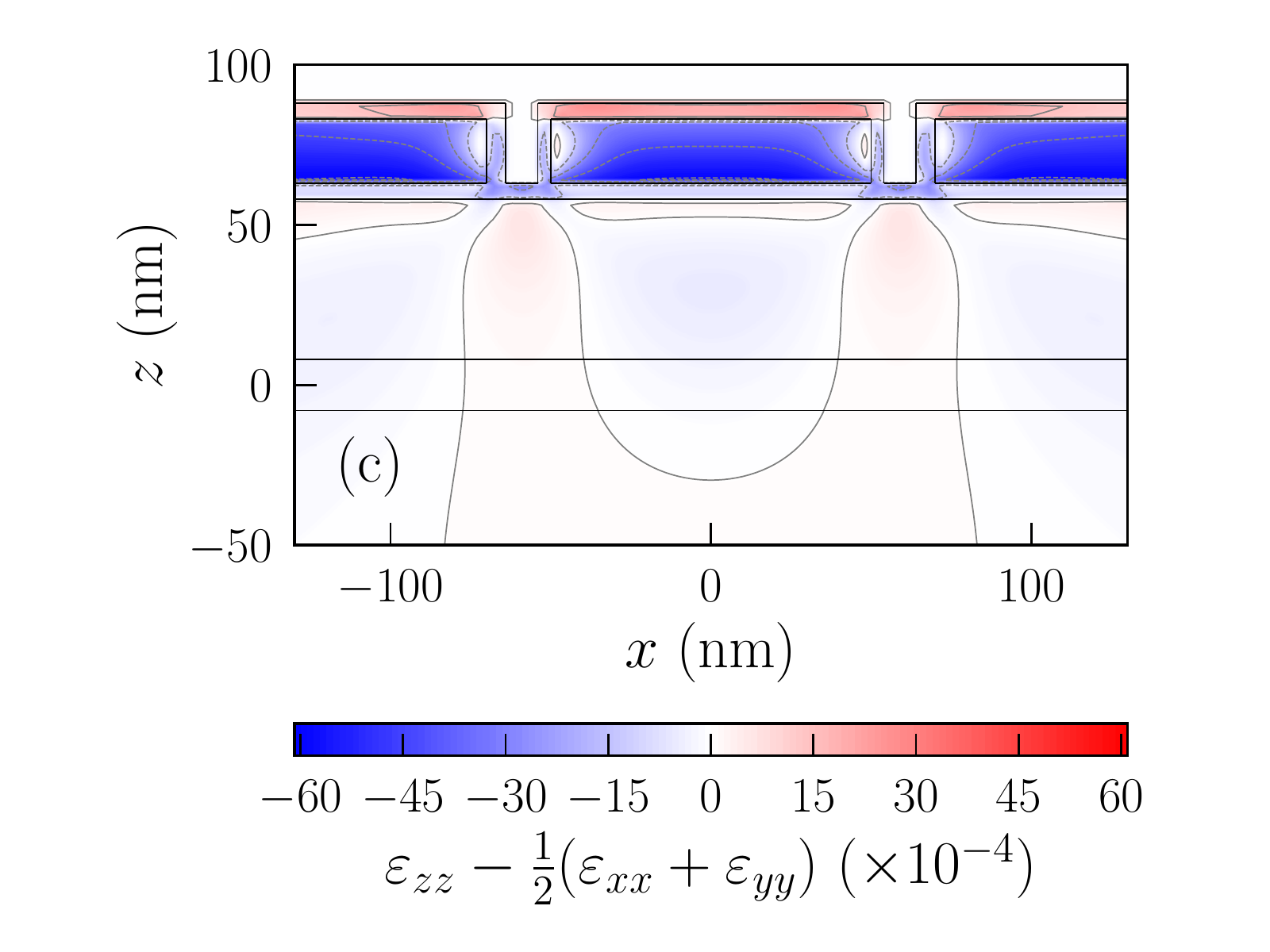}
\includegraphics[width=.45\columnwidth]{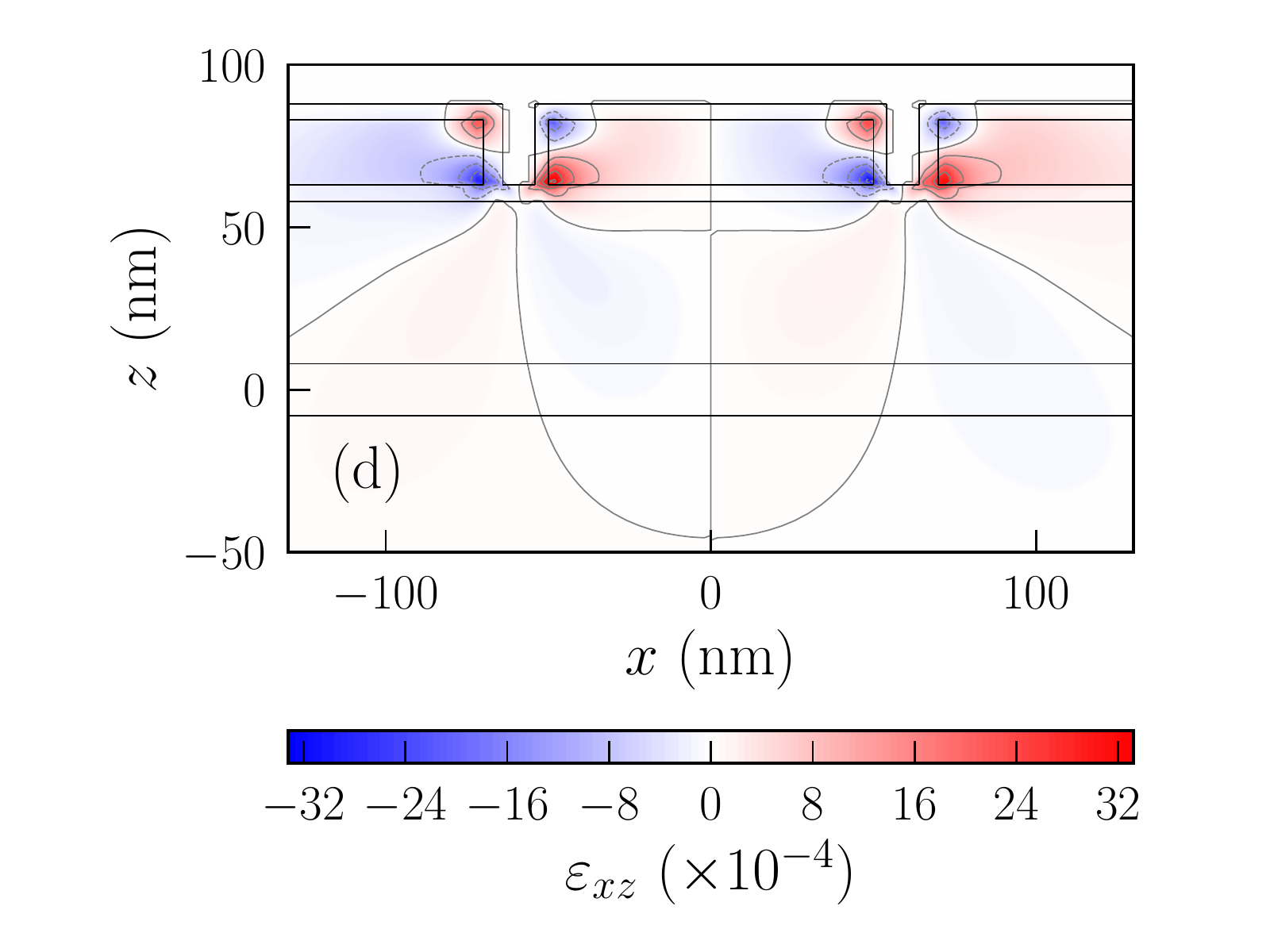}
\caption{Difference between inhomogeneous (with TC) and biaxial strains, in the $xz$ plane at $y=0$ (the vertical symmetry plane of the device). The hydrostatic strain $\delta\Omega/\Omega=\varepsilon_{xx}+\varepsilon_{yy}+\varepsilon_{zz}$ is the local, relative variation of the volume of the material. The black lines delineate the different materials.}
\label{fig:strainsxz}
\end{figure}

\section{Strains}
\label{sec:strains}

The strains in the device are computed with a 3D rectangular finite-elements method (same tensor product grid as for the finite-difference solution of the Luttinger-Kohn equations). The elastic energy density
\begin{equation}
U_\mathrm{e}=\frac{1}{2}c_{11}\left(\varepsilon_{xx}^2+\varepsilon_{yy}^2+\varepsilon_{zz}^2\right)+c_{12}\left(\varepsilon_{yy}\varepsilon_{zz}+\varepsilon_{xx}\varepsilon_{zz}+\varepsilon_{xx}\varepsilon_{yy}\right)+2c_{44}\left(\varepsilon_{yz}^2+\varepsilon_{xz}^2+\varepsilon_{xy}^2\right)
\end{equation}
is computed from the strains
\begin{equation}
\varepsilon_{\alpha\beta}(\mathbf{r})=\frac{1}{2}\left[\frac{\partial}{\partial \beta}u_\alpha(\mathbf{r})+\frac{\partial}{\partial \alpha}u_\beta(\mathbf{r})\right]
\end{equation}
where the displacement $\mathbf{u}(\mathbf{r})$ in a given element is interpolated from the corners with piecewise-linear functions. The total elastic energy (integrated over all elements) is then minimized with respect to the displacements on the grid with a conjugate-gradients method.

\begin{figure}
\centering
\includegraphics[width=.45\columnwidth]{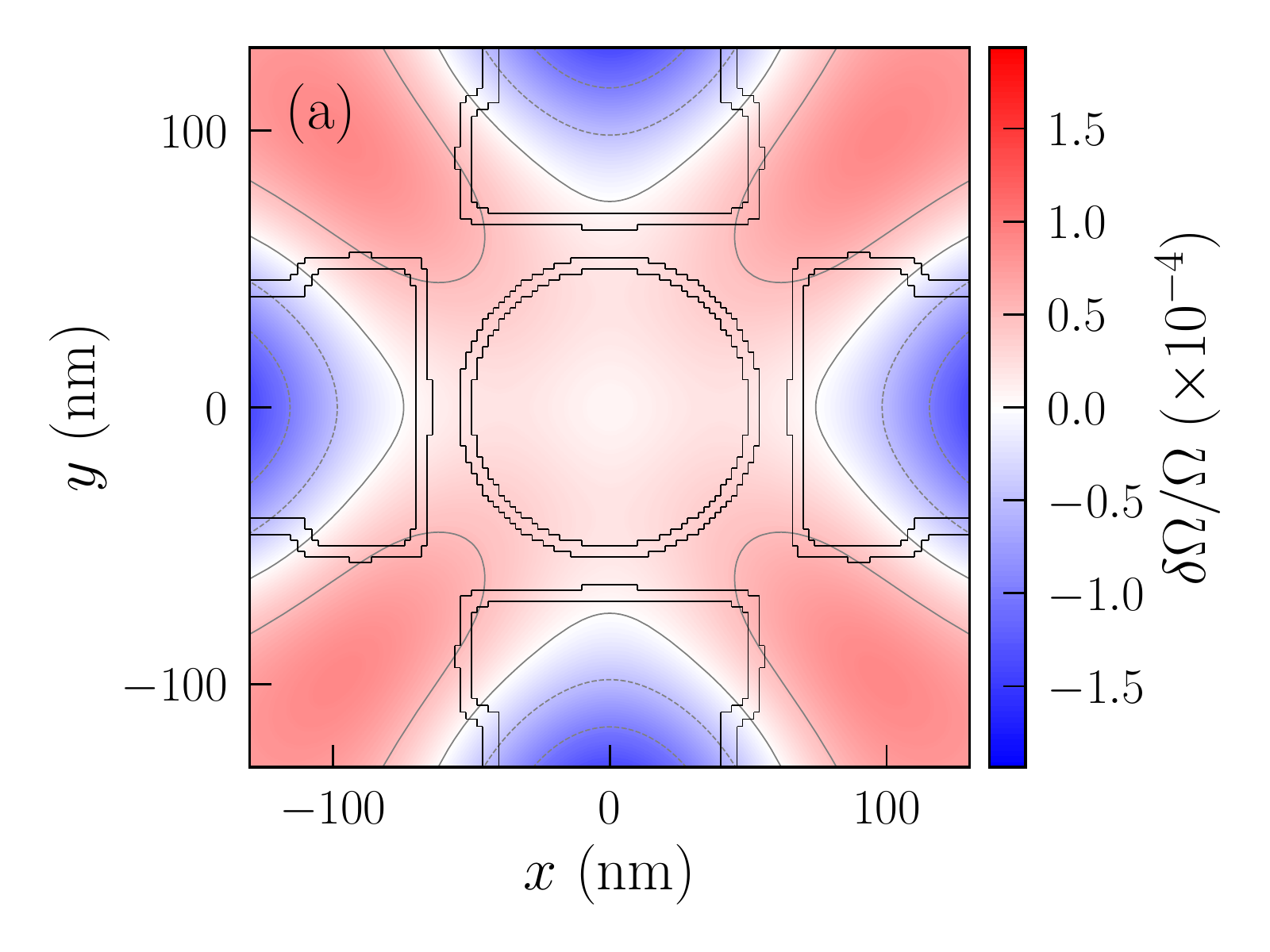}
\includegraphics[width=.45\columnwidth]{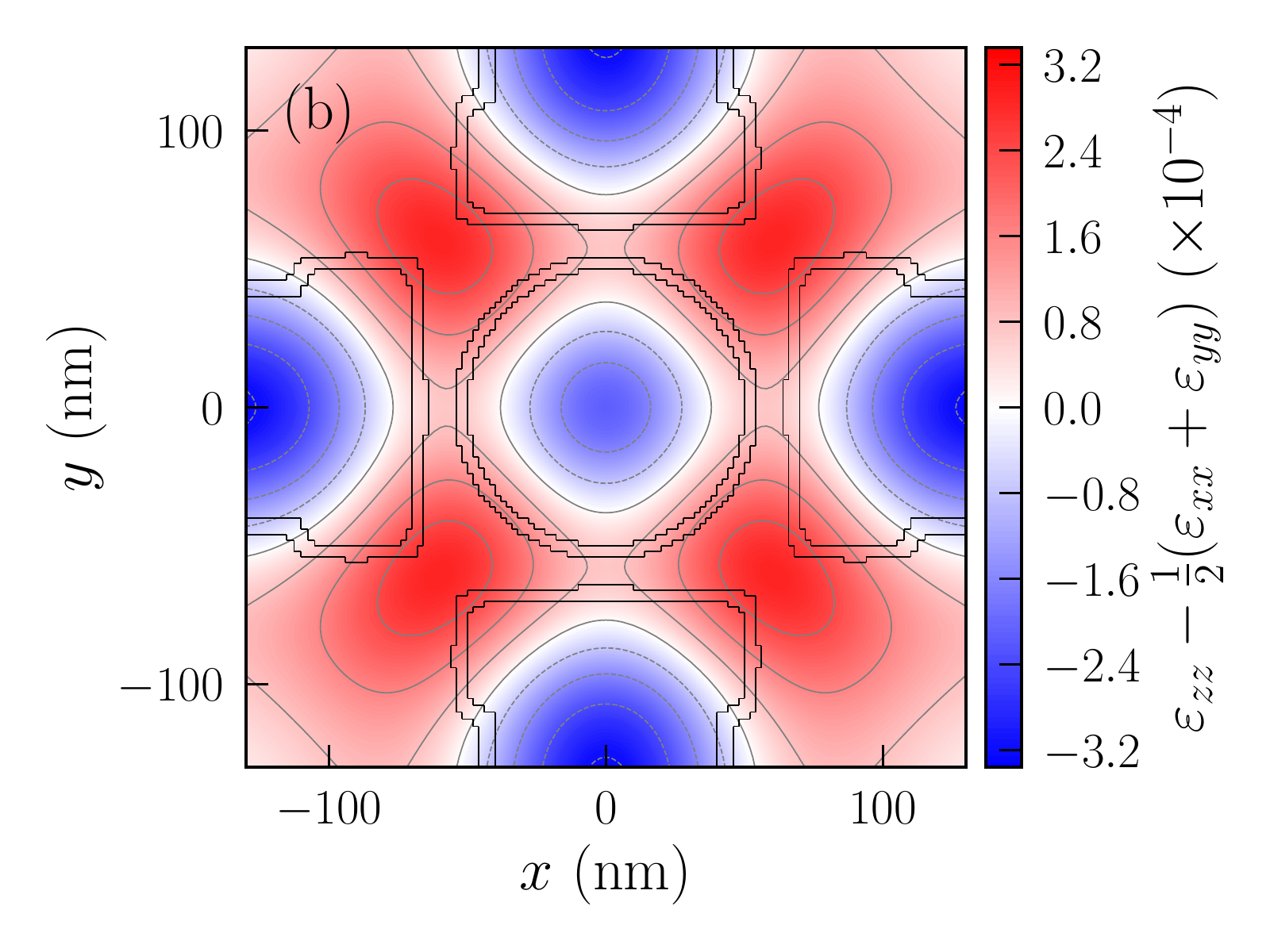} \\
\includegraphics[width=.45\columnwidth]{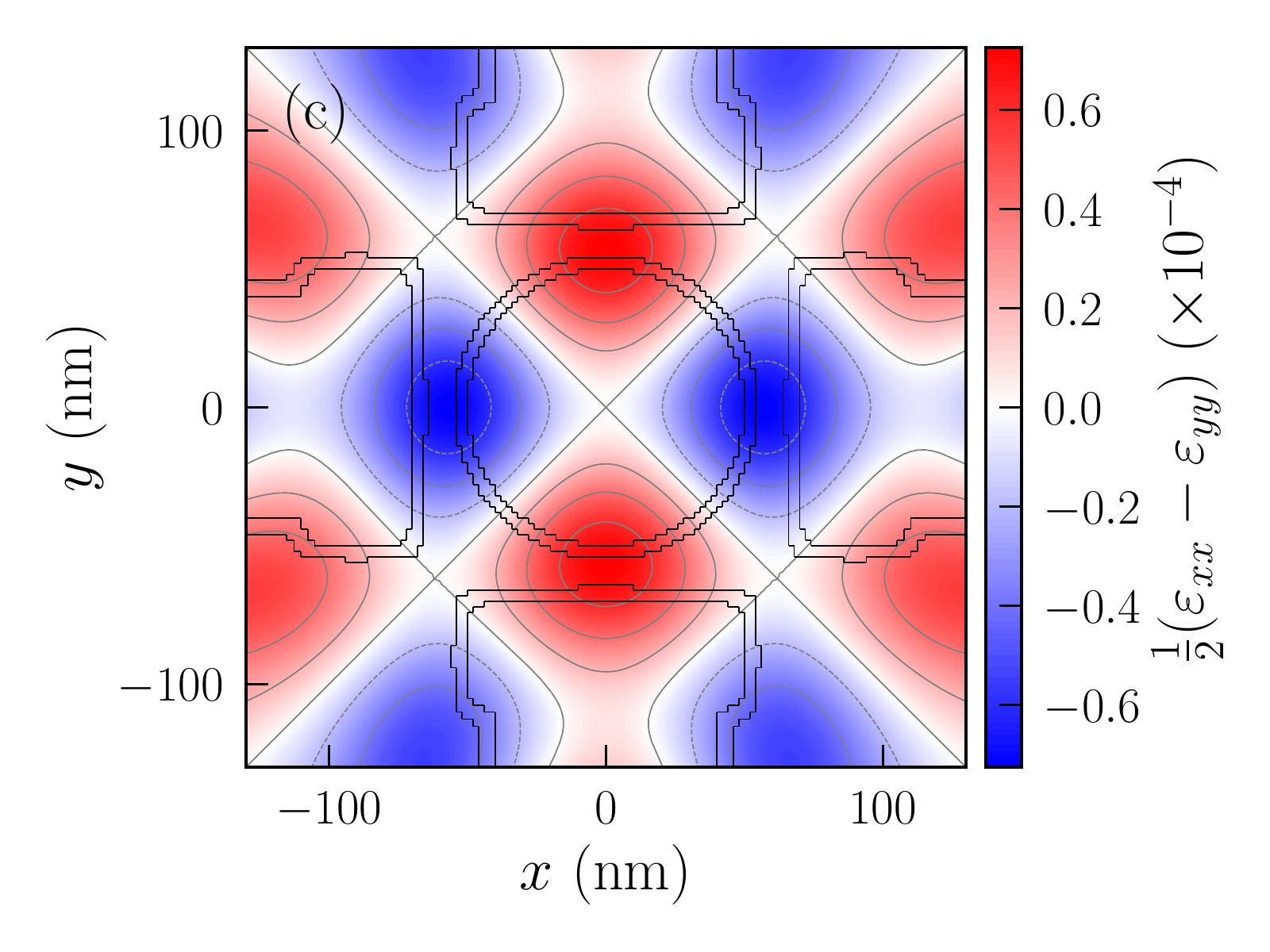}
\includegraphics[width=.45\columnwidth]{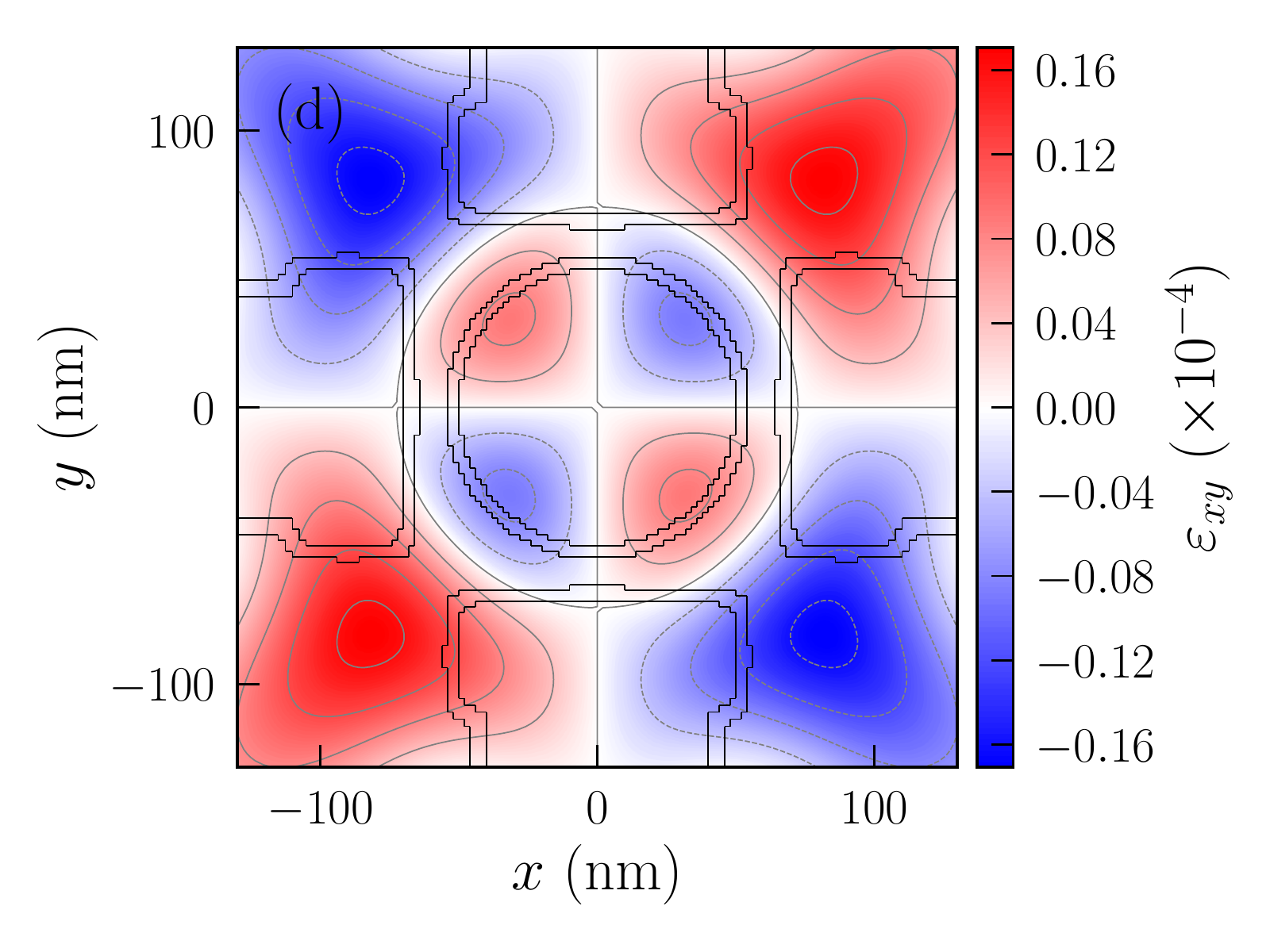} \\
\includegraphics[width=.45\columnwidth]{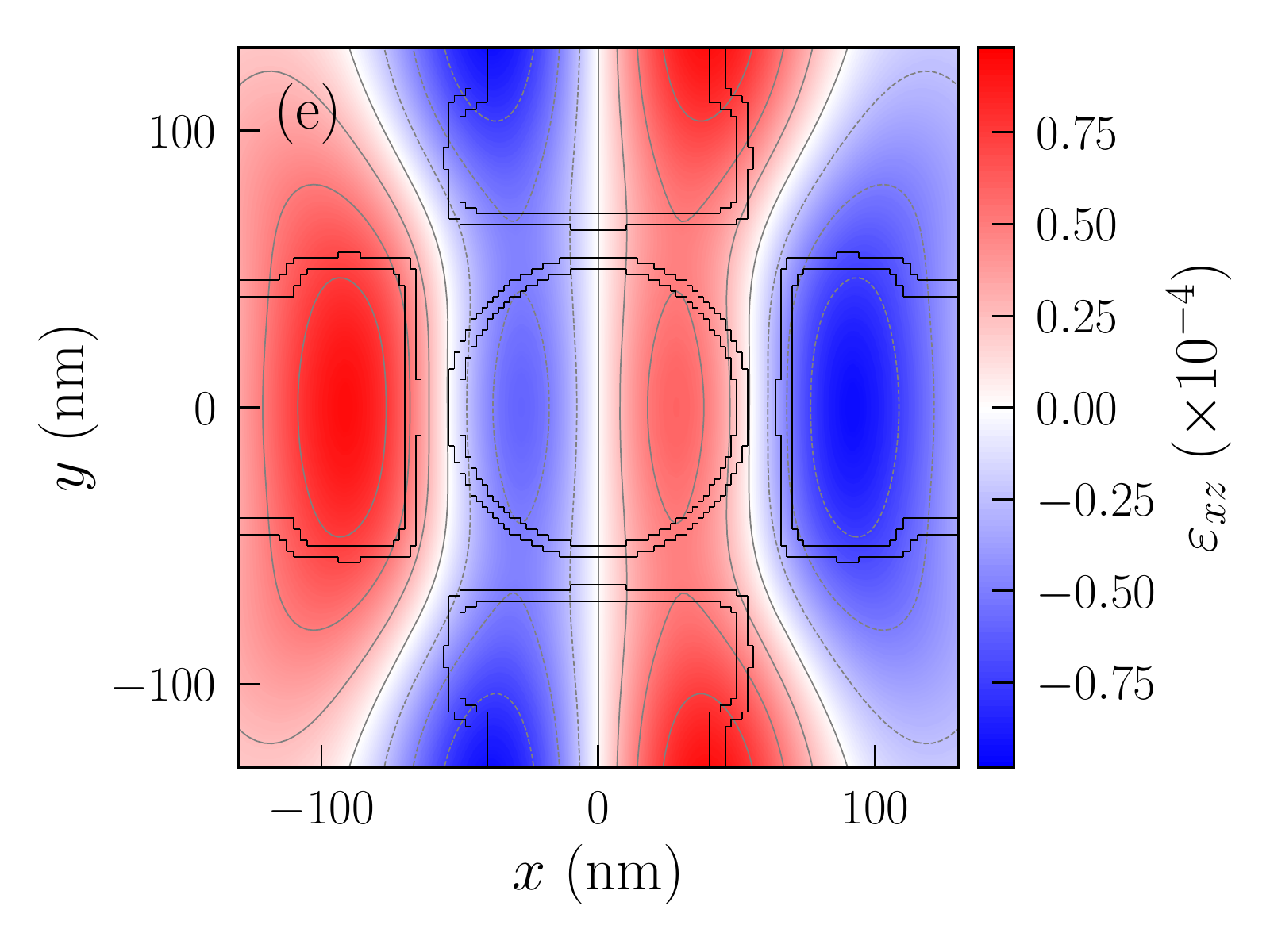}
\includegraphics[width=.45\columnwidth]{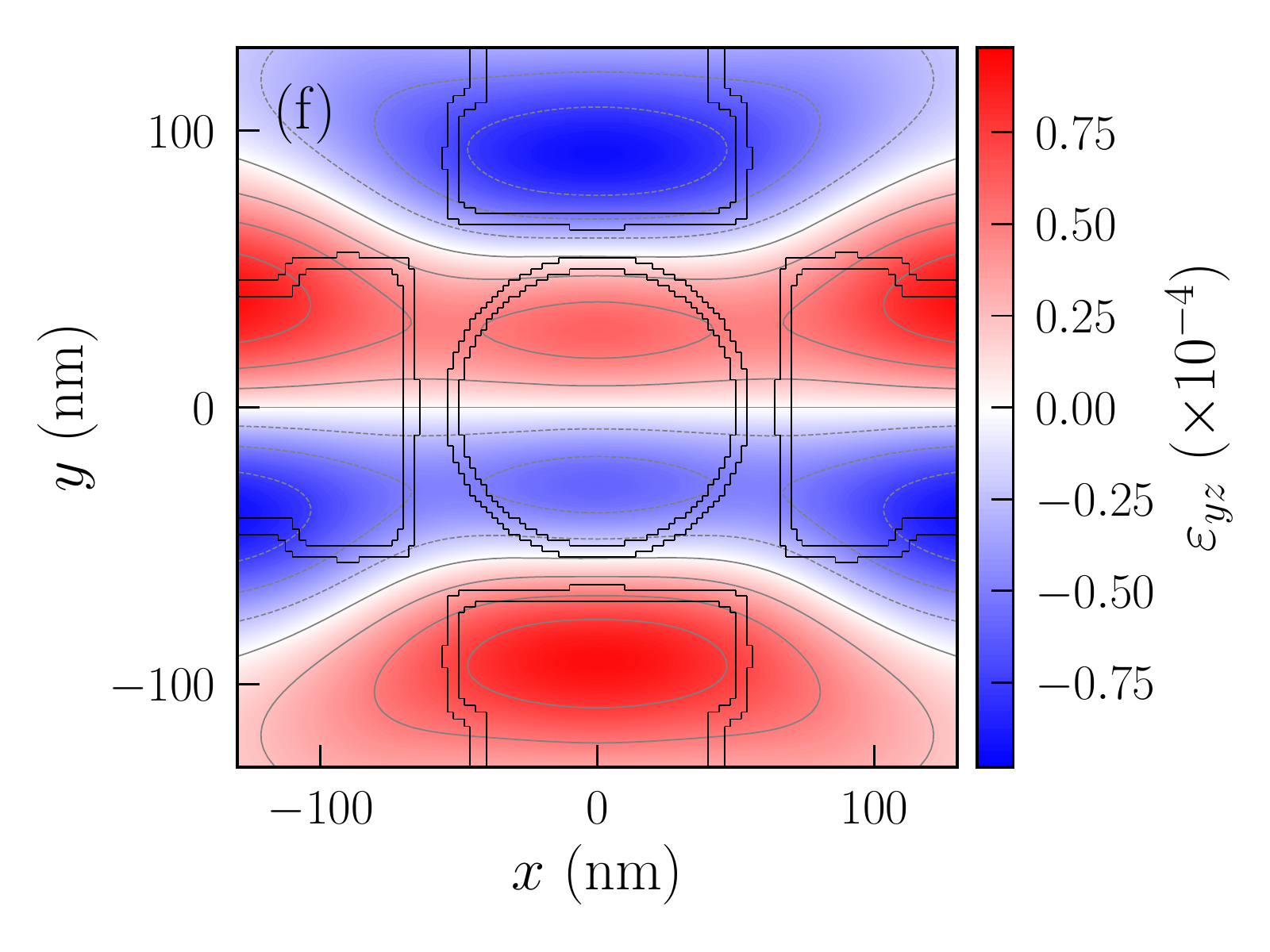} \\
\caption{Difference between inhomogeneous (with TC) and biaxial strains, in the $xy$ plane at $z=0$ (the horizontal plane through the middle of the Ge well). The hydrostatic strain $\delta\Omega/\Omega=\varepsilon_{xx}+\varepsilon_{yy}+\varepsilon_{zz}$ is the local, relative variation of the volume of the material. The black lines delineate the position of the gates (and Al$_2$O$_3$ around) at the surface of the heterostructure.}
\label{fig:strainsxy}
\end{figure}

The difference between ``inhomogeneous'' (with TC) and biaxial strains are plotted in Figs.~\ref{fig:strainsxz} and \ref{fig:strainsxy}. Figures~\ref{fig:strainsxz} and \ref{fig:strainsxy} are therefore representative of the TC strains induced by the gate stack. In the biaxial case, the residual strains in the buffer are $\varepsilon_{xx}=\varepsilon_{yy}=\varepsilon_\parallel=0.26\%$, $\varepsilon_{zz}=\varepsilon_\perp=-0.20\%$, and the strains in the Ge well are $\varepsilon_{xx}=\varepsilon_{yy}=\varepsilon_\parallel=-0.61\%$, $\varepsilon_{zz}=\varepsilon_\perp=0.45\%$ \footnote{Note that the elastic constants used in this work have been refined and are slightly different from Ref.~\cite{martinez2022hole}. Therefore, the biaxial strains in the buffer and Ge well are also slightly different, with no sizable impact on the results}.

The average in-plane strain $\varepsilon_\mathrm{p}=\tfrac{1}{2}(\varepsilon_{xx}+\varepsilon_{yy})\approx 0.27\%$ at the bottom of the central Al gate (Fig.~\ref{fig:strainsxz}a) remains close to $-\varepsilon_\mathrm{Al}=0.35\%$. This shows that the thermal contraction of the Al gate is largely hindered by the harder materials around (Al$_2$O$_3$ and Ge$_{0.8}$Si$_{0.2}$). Consequently, the TC strains induced in the Ge well below are small (of the order of $10^{-2}\%$). The magnitude of the shear strains is comparable to the hydrostatic [$\delta\Omega/\Omega=\varepsilon_{xx}+\varepsilon_{yy}+\varepsilon_{zz}$] and uniaxial [$\varepsilon_{zz}-(\varepsilon_{xx}+\varepsilon_{yy})/2)$] components.

\begin{figure}
\includegraphics[width=\textwidth]{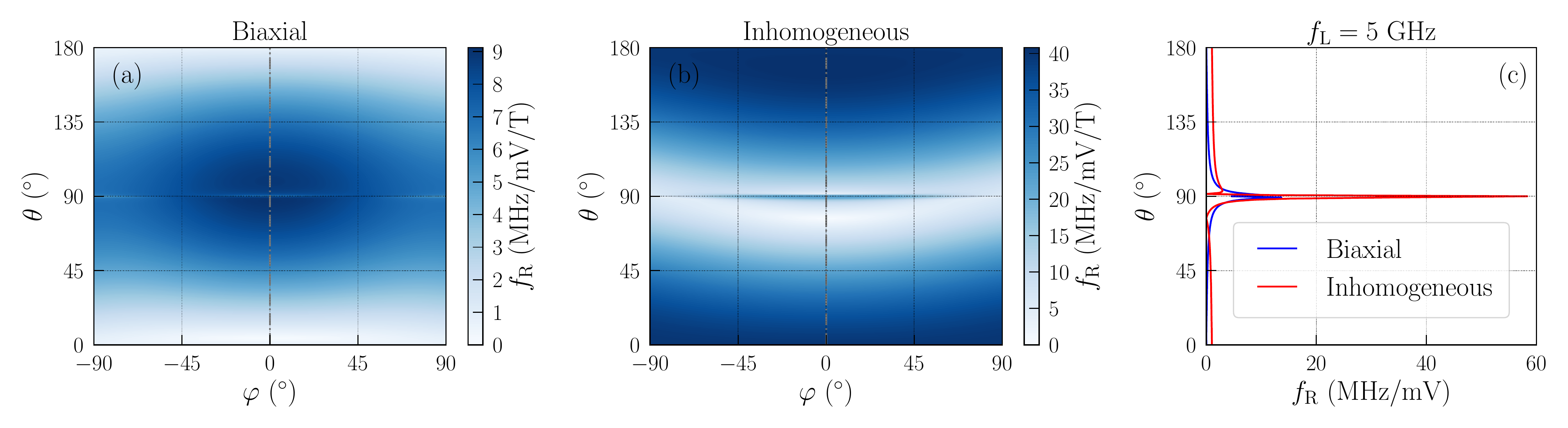}
\caption{(a, b) Maps of Rabi frequency as a function of the orientation of the magnetic field, for a drive $\delta V_\mathrm{L}=V_\mathrm{ac}\cos(2\pi f_\mathrm{L}t)$ on the L gate only ($V_\mathrm{C}=-40$\,mV). Map (a) is for homogeneous biaxial strains, and map (b) is with inhomogeneous TC strains. (c) Rabi frequency as a function of $\theta$ ($\varphi=0$) at constant Larmor frequency $f_\mathrm{L}=5$\,GHz.}
\label{fig:driveL}
\includegraphics[width=\textwidth]{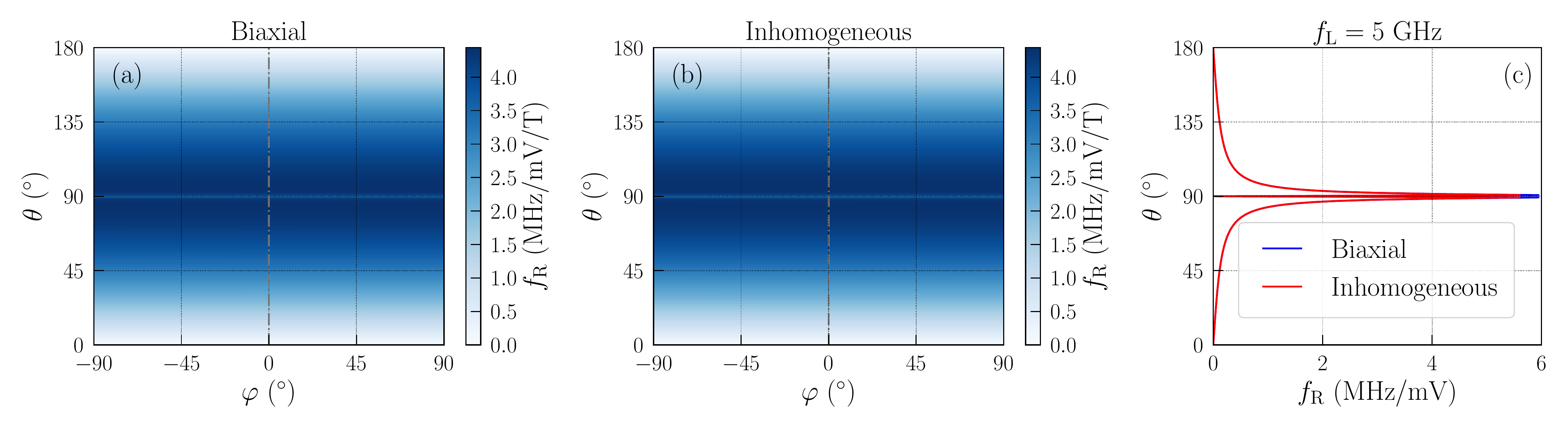}
\caption{(a, b) Maps of Rabi frequency as a function of the orientation of the magnetic field, for a drive $\delta V_\mathrm{C}=V_\mathrm{ac}\cos(2\pi f_\mathrm{L}t)$ on the C gate only ($V_\mathrm{C}=-40$\,mV). Map (a) is for homogeneous biaxial strains, and map (b) is with inhomogeneous TC strains. (c) Rabi frequency as a function of $\theta$ ($\varphi=0$) at constant Larmor frequency $f_\mathrm{L}=5$\,GHz.}
\label{fig:driveC}
\end{figure}

\YMNN{We emphasize that the C gate is the primary stressor for the dot beneath. Upon cool-down, the L/R/T/B gates actually pull in the direction opposite to the C gate and therefore decrease the shear strains in the dot. As a consequence, the Rabi frequency is slightly larger when the L/R/T/B gates are ``infinitely soft'' and do not strain the heterostructure. As an illustration, $f_\mathrm{R}$ reaches $74.4$\,MHz/mV at $f_\mathrm{L}=5$\,GHz with soft side gates (magnetic field $\mathbf{B}\parallel\mathbf{x}$), instead of $f_\mathrm{R}=58.2$\,MHz/mV on Fig. 3c of the main text.}

\section{Driving with the L or C gate only}
\label{sec:LCdrive}

The maps of Rabi frequency for a drive $\delta V_\mathrm{L}=V_\mathrm{ac}\cos(2\pi f_\mathrm{L}t)$ on the L gate only are plotted in Fig.~\ref{fig:driveL}. In biaxial strains, the motion of the dot in the non-separable potential of the gates results in a non-zero $\gt_{zx}^\prime$, and the cubic Rashba spin-orbit interaction (SOI) in a non-zero $\gt_{xz}^\prime$, as in Fig.~3 of the main text. These contributions are, however, outweighed by direct modulations of the principal $\gt$-factors $\gt_{xx}$, $\gt_{yy}$ and $\gt_{zz}$ by the inhomogeneous AC electric field of the L gate that squeezes the dot dynamically (see Table~\ref{tab:symmetries}) \cite{martinez2022hole}. These modulations give rise to the broad feature that differentiates Fig.~\ref{fig:driveL} from Fig.~3. 

When TC is accounted for, these mechanisms are superseded by the same strain-induced modulations of $\gt_{zx}$ and $\gt_{xz}$ as in the main text [Eqs.~(7c) and (8)]. At variance with the out-of-phase L/R drive of Fig.~3, the deformations of the dot in the inhomogeneous strains also gives rise to finite $\gt_{xx}^\prime$, $\gt_{yy}^\prime$ and $\gt_{zz}^\prime$. They are however much smaller than $\gt_{zx}^\prime$ and $\gt_{xz}^\prime$, so that the anisotropy and magnitude of the Rabi frequency are comparable to Fig.~3.

The maps of Rabi frequency for a drive $\delta V_\mathrm{C}=V_\mathrm{ac}\cos(2\pi f_\mathrm{L}t)$ on the C gate only are plotted in Fig.~\ref{fig:driveC}. For symmetry reasons, such a drive can only modulate the principal $\gt$-factors $\gt_{xx}$, $\gt_{yy}$ and $\gt_{zz}$ as the dot ``breathes'' in the AC electric field ($\gt_{xz}^\prime=\gt_{xz}^\prime=0$, see Table~\ref{tab:symmetries}) \cite{martinez2022hole}. However, the first-order contributions of strains to $\gt_{xx}^\prime$ and $\gt_{yy}^\prime$ are zero given the $\varepsilon_{\alpha\beta}$'s shown in Fig.~\ref{fig:strainsxy}. The TC strains only give rise to second-order variations of the $\gt$-factors, in particular through modulations of the heavy-hole/light-hole gap $\Delta_\mathrm{LH}$. As a consequence, the Rabi frequencies are almost the same with and without TC strains. They are, in particular, zero for in-plane magnetic fields, at variance with the previous cases \cite{martinez2022hole}.

\section{Rabi frequencies for a thinner Ge$_{0.8}$Si$_{0.2}$ barrier}
\label{sec:barrier}

\begin{figure}
\includegraphics[width=\textwidth]{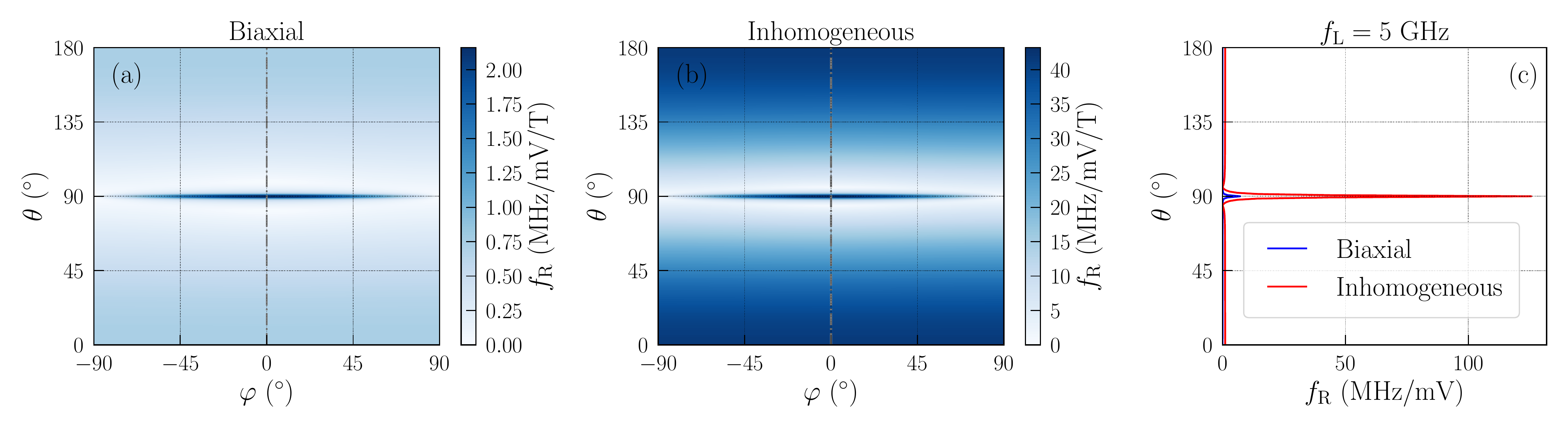}
\caption{(a, b) Maps of Rabi frequency as a function of the orientation of the magnetic field, for opposite drives $\delta V_\mathrm{L}=-\delta V_\mathrm{R}=\tfrac{1}{2}V_\mathrm{ac}\cos(2\pi f_\mathrm{L}t)$ on the L and R gates ($V_\mathrm{C}=-40$\,mV). The top Ge$_{0.8}$Si$_{0.2}$ barrier is 25 nm thick. Map (a) is for homogeneous biaxial strains, and map (b) is with inhomogeneous TC strains. (c) Rabi frequency as a function of $\theta$ ($\varphi=0$) at constant Larmor frequency $f_\mathrm{L}=5$\,GHz.}
\label{fig:thin}
\vspace{.5cm}
(a)
\begin{tabular}{|c|c|c|} \hline
 & $\sigma_{yz}$ & $\sigma_{xz}$ \\ \hline
$\mathbf{E}_\mathrm{ac}$ even &  
$\begin{pmatrix}
\bullet & 0 & 0 \\
0 & \bullet & \bullet \\
0 & \bullet & \bullet
\end{pmatrix}$ &
$\begin{pmatrix}
\bullet & 0 & \bullet \\
0 & \bullet & 0 \\
\bullet & 0 & \bullet
\end{pmatrix}$ \\ \hline
$\mathbf{E}_\mathrm{ac}$ odd & 
$\begin{pmatrix}
0 & \bullet & \bullet \\
\bullet & 0  & 0 \\
\bullet & 0 & 0
\end{pmatrix}$ & 
$\begin{pmatrix}
0 & \bullet & 0 \\
\bullet & 0 & \bullet \\
0 & \bullet & 0
\end{pmatrix}$  \\ \hline
Other & 
$\begin{pmatrix}
\bullet & \bullet & \bullet \\
\bullet & \bullet & \bullet \\
\bullet & \bullet & \bullet
\end{pmatrix}$ & 
$\begin{pmatrix}
\bullet & \bullet & \bullet \\
\bullet & \bullet & \bullet \\
\bullet & \bullet & \bullet
\end{pmatrix}$ \\
\hline
\end{tabular}
\ \ \ \ \ (b)
\begin{tabular}{|c|c|c|c|} \hline
\multirow{2}{*}{Drive} & Parity of $\mathbf{E}_\mathrm{ac}$ & Parity of $\mathbf{E}_\mathrm{ac}$ & \multirow{2}{*}{$\gt^\prime$} \\
 & wrt $\sigma_{yz}$ & wrt $\sigma_{xz}$ & \\
\hline
Opposite L/R & Odd & Even & $\begin{pmatrix}
0 & 0 & \bullet \\
0 & 0  & 0 \\
\bullet & 0 & 0
\end{pmatrix}$ \\ \hline
L & None & Even & $\begin{pmatrix}
\bullet & 0 & \bullet \\
0 & \bullet & 0 \\
\bullet & 0 & \bullet
\end{pmatrix}$ \\ \hline
C & Even & Even & $\begin{pmatrix}
\bullet & 0 & 0 \\
0 & \bullet & 0 \\
0 & 0 & \bullet
\end{pmatrix}$ \\
\hline
\end{tabular}
\captionof{table}{(a) Constraints on the shape of $\gt^\prime$ set by the mirror planes $\sigma_{yz}$ and $\sigma_{xz}$ of the device of Fig.~1, depending whether the AC electric field $\mathbf{E}_\mathrm{ac}$ is even [$\mathbf{E}_\mathrm{ac}(\sigma_{\alpha\beta}(\mathbf{r}))=\sigma_{\alpha\beta}(\mathbf{E}_\mathrm{ac}(\mathbf{r}))$], odd [$\mathbf{E}_\mathrm{ac}(\sigma_{\alpha\beta}(\mathbf{r}))=-\sigma_{\alpha\beta}(\mathbf{E}_\mathrm{ac}(\mathbf{r}))$], or does not show any relevant parity under that mirror transformation. The black dots are the non-zero matrix elements \cite{Venitucci18}. (b) Shape of $\gt^\prime$ set by symmetries for the different drives considered in this work: opposite drives on the L and R gates, drive on the L gate only, and on the C gate only. The second and third columns are the parities of $\mathbf{E}_\mathrm{ac}$ with respect to the $\sigma_{yz}$ and $\sigma_{xz}$ mirrors. The last column is the shape of the $\gt^\prime$ constructed from the intersection of the relevant patterns of Table (a).}
\label{tab:symmetries}
\end{figure}

The maps of Rabi frequencies computed for a 25 nm thick upper Ge$_{0.8}$Si$_{0.2}$ barrier are plotted in Fig.~\ref{fig:thin}. The dot is driven with opposite modulations $\delta V_\mathrm{L}=-\delta V_\mathrm{R}=\tfrac{1}{2}V_\mathrm{ac}\cos(2\pi f_\mathrm{L}t)$ on the L and R gates, as in the main text. The distribution of TC strains in the substrate is little affected by this change, the elastic constants of Ge and Ge$_{0.8}$Si$_{0.2}$ being very close. However, the TC strains in the Ge well are much greater, since the latter is brought closer to the Al gates. In particular, the shear strains $\varepsilon_{xz}$ and $\varepsilon_{yz}$ in the Ge well are about twice larger than for a 50 nm thick barrier. The calculated Rabi frequencies for in-plane magnetic fields are, therefore, enhanced by a factor $\approx 2.2$ (also when the dot is driven with the L gate only).

\section{Rabi frequencies as a function of metal and oxide thicknesses}
\label{sec:metal}

\begin{figure}
\includegraphics[width=.475\textwidth]{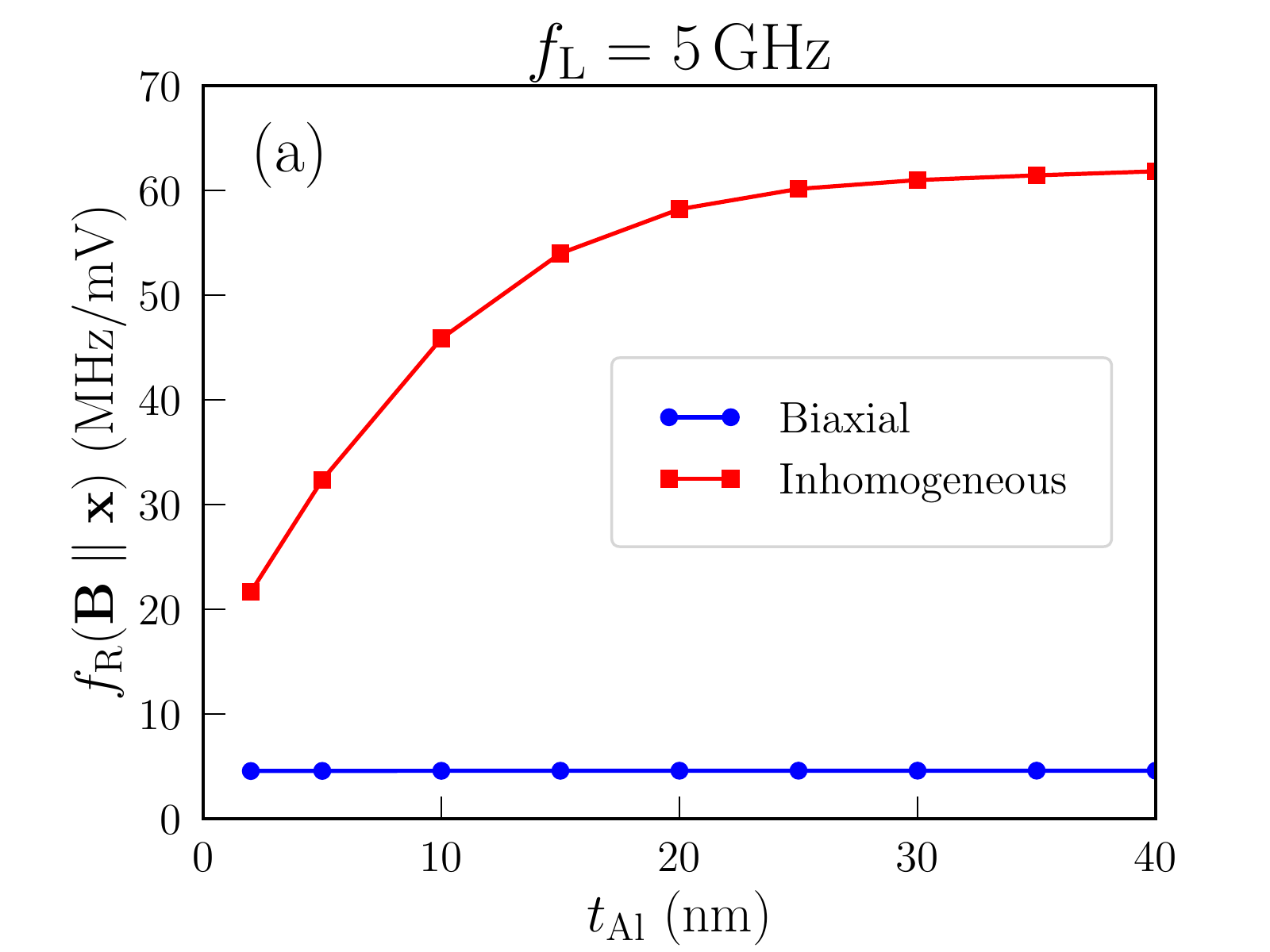}
\includegraphics[width=.475\textwidth]{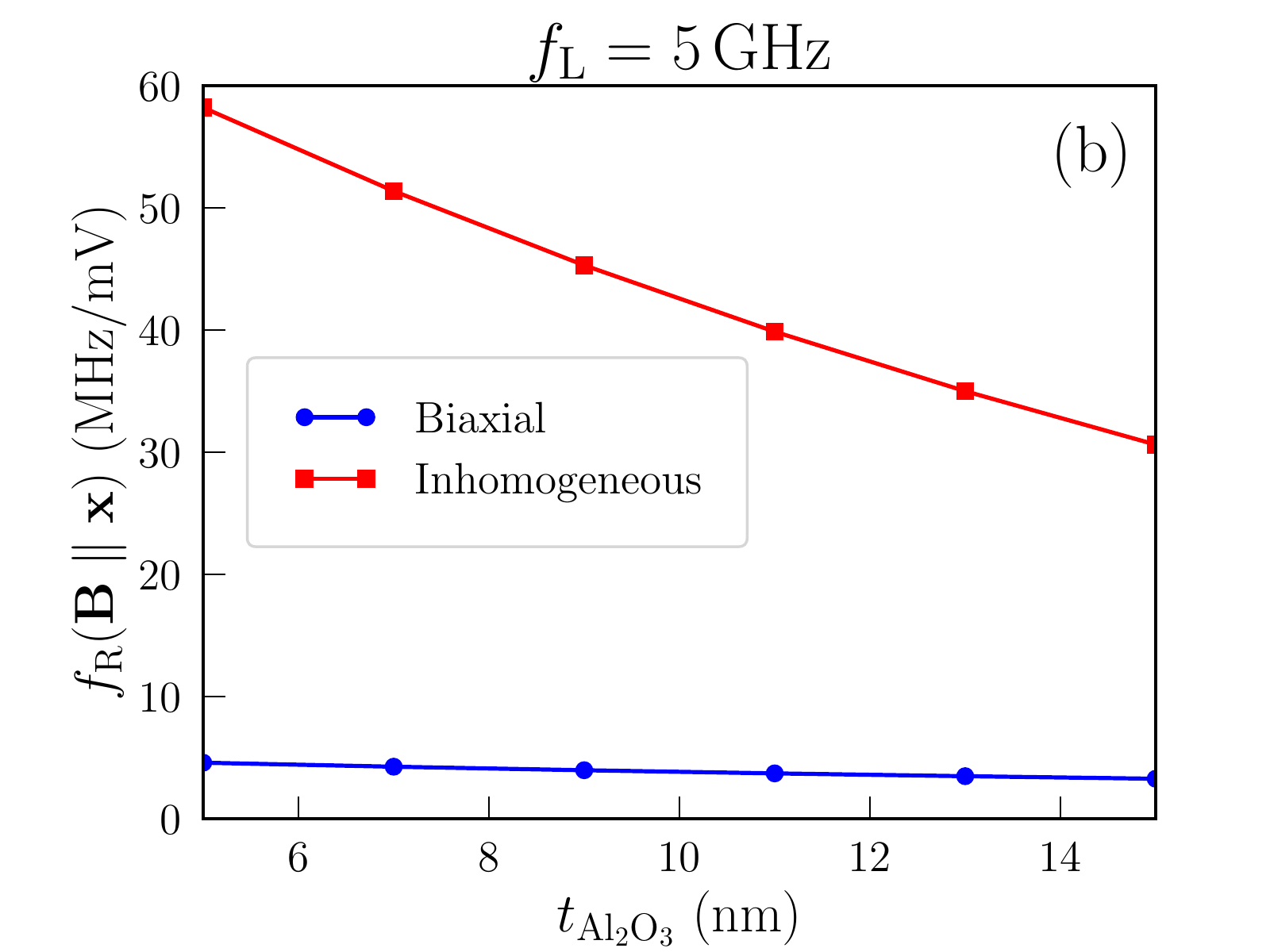}
\caption{\YMN{Rabi frequency $f_\mathrm{R}(\mathbf{B}\parallel\mathbf{x})$ at constant Larmor frequency $f_\mathrm{L}=5$\,GHz as a function of (a) the thickness $t_\mathrm{Al}$ of the metal gates and (b) the thickness $t_\mathrm{Al_2O_3}$ of the Al$_2$O$_3$ layer between the gates and the heterostructure. The hole is driven with opposite modulations $\delta V_\mathrm{L}=-\delta V_\mathrm{R}=\tfrac{1}{2}V_\mathrm{ac}\cos(2\pi f_\mathrm{L}t)$ on the L and R gates ($V_\mathrm{C}=-40$\,mV). The data are plotted in biaxial and with inhomogeneous TC strains. The top Ge$_{0.8}$Si$_{0.2}$ barrier is 50 nm thick; $t_\mathrm{Al_2O_3}=5$\,nm in (a) and $t_\mathrm{Al}=20$\,nm in (b).}}
\label{fig:thickness}
\end{figure}

\YMN{The strain-induced SOI in a heterostructure can be most easily probed by changing either the nature or the thickness of the main stressors, namely the metal gates. The Rabi frequency $f_\mathrm{R}(\mathbf{B}\parallel\mathbf{x})$ computed at constant Larmor frequency $f_\mathrm{L}=5$\,GHz is plotted as a function of the gate thickness $t_\mathrm{Al}$ in Fig. \ref{fig:thickness}a (opposite modulations on the L/R gates). In homogeneous biaxial strains, increasing the metal thickness has little effect on the electrostatics of the deeply buried well (the Rabi frequency increases from $f_\mathrm{R}=4.566$\,MHz/mV for $t_\mathrm{Al}=5$\,nm to $f_\mathrm{R}=4.590$\,MHz/mV for $t_\mathrm{Al}=40$\,nm). When inhomogeneous TC strains are accounted for, the Rabi frequency shows a much stronger dependence on $t_\mathrm{Al}$. It increases rapidly for small metal thickness then saturates once $t_\mathrm{Al}$ is a significant fraction of the metal line width so that stress can be relieved through side facets deformation. Note that the Rabi frequencies with and without TC strains do not tend to the same limits when $t_\mathrm{Al}\to0$ due to the residual stress imposed by the aluminium oxide.}

\YMN{Actually, Al$_2$O$_3$ is a rather hard gate oxide, whose dimensions can have a significant impact on both strains and electrostatics. The Rabi frequency  $f_\mathrm{R}(\mathbf{B}\parallel\mathbf{x})$ is likewise plotted as a function of the thickness $t_\mathrm{Al_2O_3}$ of the bottom Al$_2$O$_3$ layer between the gates and heterostructure. In biaxial strains, the Rabi oscillations slow down when increasing $t_\mathrm{Al_2O_3}$ due to the loss of electrostatic control ($\Delta f_\mathrm{R}/f_\mathrm{R}=-29$\% from $f_\mathrm{R}=4.585$\,MHz/mV for $t_\mathrm{Al_2O_3}=5$\,nm to $f_\mathrm{R}=3.272$\,MHz/mV for $t_\mathrm{Al_2O_3}=15$\,nm). The decrease is much faster with TC strains as Al$_2$O$_3$ also limits the contraction of the metal gates, hence the cool-down strains transferred to the heterostructure. However, the relative decrease $\Delta f_\mathrm{R}/f_\mathrm{R}=-47$\% is only $1.65\times$ larger than in biaxial strains. The dependence of the Rabi frequency on $t_\mathrm{Al_2O_3}$ is not, therefore, as conclusive as its dependence on $t_\mathrm{Al}$ as to the prevalence of strain-induced spin-orbit interactions.}

\section{Full set of strain-induced spin-orbit interactions}
\label{sec:deltags}

The interplay between $R_\mathrm{K}$ or $S_\mathrm{K}$ and the strain Hamiltonian $H_\varepsilon$ gives rise to following linear-in-momentum spin-orbit interactions in the $\{\ket{+\tfrac{3}{2}},\,\ket{-\tfrac{3}{2}}\}$ basis set (the counterparts of Eq.~(8) of the main text):
\begin{equation}
\delta{\cal H}_\mathrm{so}=\frac{\hbar}{m_0}\sum_{\alpha,\,\beta\in\{x,\,y,\,z\}}\left[c_{\alpha\beta}p_\alpha-i\frac{\hbar}{2}\left(\frac{\partial c_{\alpha\beta}}{\partial\alpha}\right)\right]\sigma_\beta\,,
\label{eq:rashba2}
\end{equation}
where:
\begin{subequations}
\label{eq:calphabeta}
\begin{align}
c_{xx}&=\frac{\sqrt{3}d_{v}}{\Delta_\mathrm{LH}}\left[\gamma_{2}\deyzdx+\gamma_{3}\dexzdy-\gamma_{3}\dexydz\right] \\
c_{xy}&=\frac{1}{2\Delta_\mathrm{LH}}\left[+3b_{v}\gamma_{3}\dexxeyydz+2\sqrt{3}d_{v}\left(\gamma_{3}\deyzdy-\gamma_{2}\dexzdx\right)\right] \\
c_{xz}&=\frac{1}{2\Delta_\mathrm{LH}}\left[-3b_{v}\gamma_{3}\dexxeyydy+2\sqrt{3}d_{v}\left(\gamma_{3}\deyzdz+\gamma_{2}\dexydx\right)\right] \\
c_{yx}&=\frac{1}{2\Delta_\mathrm{LH}}\left[-3b_{v}\gamma_{3}\dexxeyydz+2\sqrt{3}d_{v}\left(\gamma_{3}\dexzdx-\gamma_{2}\deyzdy\right)\right] \\
c_{yy}&=\frac{\sqrt{3}d_{v}}{\Delta_\mathrm{LH}} \left[\gamma_{2}\dexzdy+\gamma_{3}\deyzdx-\gamma_{3}\dexydz\right] \\
c_{yz}&=\frac{1}{2\Delta_\mathrm{LH}}\left[-3b_{v}\gamma_{3}\dexxeyydx-2\sqrt{3}d_{v}\left(\gamma_{3}\dexzdz+\gamma_{2}\dexydy\right)\right]
\end{align}
\begin{align}
c_{zx}&=\frac{\gamma_3}{2\Delta_\mathrm{LH}}\left[-3b_{v}\dexxeyydy-2\sqrt{3}d_{v}\dexydx\right] \\
c_{zy}&=\frac{\gamma_3}{2\Delta_\mathrm{LH}}\left[+3b_{v}\dexxeyydx-2\sqrt{3}d_{v}\dexydy\right] \\
c_{zz}&=\frac{\sqrt{3}d_{v}\gamma_3}{\Delta_\mathrm{LH}}\left[\deyzdx-\dexzdy\right]\,.
\end{align}
\end{subequations}
These interactions couple the spin to the momentum of the hole in the strain gradients that act as an effective electric field. In general, the $c_{\alpha\beta}$'s (or equivalently the generalized spin-orbit lengths $\ell_\mathrm{so}^{\alpha\beta}\propto c_{\alpha\beta}^{-1}$) are spatially dependent \cite{Dolcini18}. The hermiticity of $\delta{\cal H}_\mathrm{so}$ in such an inhomogeneous SOI is ensured by the $\tfrac{\partial}{\partial\alpha}c_{\alpha\beta}$ term of Eq.~(\ref{eq:rashba2}). Inhomogeneous $c_{\alpha\beta}$'s also result in a coupling between the orbital motion of the hole in the magnetic vector potential and the position of the dot (when substituting $p_\alpha\to -i\hbar\tfrac{\partial}{\partial\alpha}+eA_\alpha$), which contributes to the Rabi oscillations (see section \ref{sec:rashba}).

\begin{figure}
\centering
\includegraphics[width=.45\columnwidth]{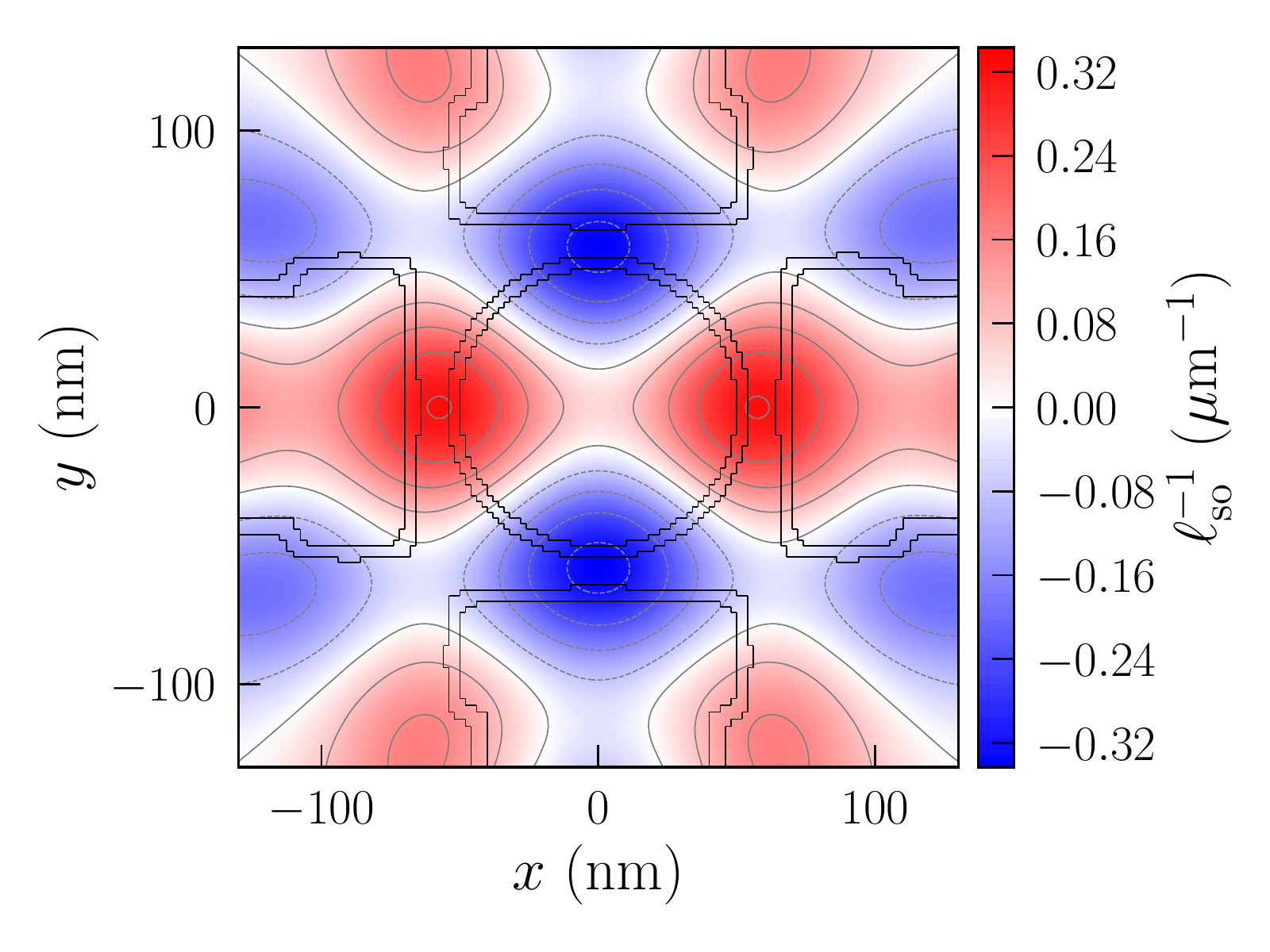}
\caption{The inverse spin-orbit length $\ell_\mathrm{so}^{-1}$ in the $xy$ plane at $z=0$ (the horizontal plane through the middle of the Ge well). This quantity, as defined given by Eq.~(8) of the main text, is signed and is strongly inhomogeneous. The black lines delineate the position of the gates (and Al$_2$O$_3$ around) at the surface of the heterostructure.}
\label{fig:lsoxy}
\end{figure}

In the setup of the main text, the dominant interactions are the $\propto p_x\sigma_y$ and $\propto p_y\sigma_x$ terms, which are mostly induced by the shear strain gradients $\tfrac{\partial}{\partial x}\varepsilon_{xz}$ and $\tfrac{\partial}{\partial y}\varepsilon_{yz}$. The inverse spin-orbit length $\ell_\mathrm{so}^{-1}$ as defined by Eq.~(8) is plotted in Fig.~\ref{fig:lsoxy}. It is, as discussed above, inhomogeneous and signed. The spin-orbit lengths remain however too long to be efficiently exploited at small Larmor frequencies, as shown by Fig.~3c. \YMNN{We also emphasize that the average $\ell_\mathrm{so}^{-1}$ is zero along the way between two identical dots with the same strains. This shall limit the contribution of strained-induced SOI to the spin-flip tunneling between neighboring dots that complicates the management of exchange interactions and is responsible for leakage in the spin-blockade regime \cite{Hung17}}.

\section{Effect of inhomogeneous strains on the coherence}
\label{sec:coherence}

\YMN{Spin-orbit interactions -- whatever their nature -- couple the spin to electric fields. They thus allow for electrical manipulation but promote dephasing, the most limiting decoherence mechanism in spin qubits.}

\YMN{As discussed in Ref.~\cite{Piot22}, the sensitivity of a spin to electrical noise can be generally characterized by the ``longitudinal electric spin susceptibilities'':
\begin{equation}
\mathrm{LSES}(A_\mathrm{n})=\frac{\partial f_\mathrm{L}}{\partial A_\mathrm{n}}
\end{equation}
where $A_\mathrm{n}$ is some fluctuating parameter that modulates the Larmor frequency. For $1/f$ charge noise in particular, the coherence decays as $\exp[-(t/T_2^*)^2]$ where $\Gamma_2^*=1/T_2^*=\sqrt{2}\pi\delta A_\mathrm{n}^\mathrm{rms}|\mathrm{LSES}(A_\mathrm{n})|$ \cite{Piot22} and $\delta A_\mathrm{n}^\mathrm{rms}$ are the rms fluctuations of $A_\mathrm{n}$. Since both the LSES and the Rabi frequency are, to first-order, proportional to the spin-orbit coupling strength, enhancing the latter does not necessarily degrade, on average, the quality factor $Q_2^*=2f_\mathrm{R}T_2^*$ (the number of $\pi$ rotations that can be achieved within $T_2^*$). There may, moreover, be ``sweet spots'' or even ''sweet lines'' as a function of the orientation of the magnetic field where the relevant LSESs are zero and the qubit is decoupled (to first-order) from electrical noise. The maximum Rabi frequency usually lies on such a sweet line owing to reciprocal sweetness relations between longitudinal and transverse spin electric susceptibilites \cite{michal2022tunable}. At these particular magnetic field orientation(s), the Rabi frequency and the coherence times $T_2^*$ are both optimal.}

\YMN{We can take $A_\mathrm{n}\equiv V_\mathrm{C}$ and $A_\mathrm{n}\equiv V_\mathrm{LR}=V_\mathrm{L}-V_\mathrm{R}$ as probes of the sensitivity of the hole spin to (quasi) vertical and in-plane electric field fluctuations, respectively. We hence define the quality factors:
\begin{subequations}
\begin{align}
Q_{2\perp}^*&=\frac{\sqrt{2}}{\pi}\frac{V_\mathrm{ac}}{\delta V_\mathrm{C}^\mathrm{rms}}f_\mathrm{R}\left|\frac{\partial f_\mathrm{L}}{\partial V_\mathrm{C}}\right|^{-1}=\frac{V_\mathrm{ac}}{\delta V_\mathrm{C}^\mathrm{rms}}\hat{Q}_{2\perp}^*\text{\ with }\hat{Q}_{2\perp}^*=\frac{\sqrt{2}}{\pi}f_\mathrm{R}\left|\frac{\partial f_\mathrm{L}}{\partial V_\mathrm{C}}\right|^{-1} \\
Q_{2\parallel}^*&=\frac{\sqrt{2}}{\pi}\frac{V_\mathrm{ac}}{\delta V_\mathrm{LR}^\mathrm{rms}}f_\mathrm{R}\left|\frac{\partial f_\mathrm{L}}{\partial V_\mathrm{LR}}\right|^{-1}=\frac{V_\mathrm{ac}}{\delta V_\mathrm{LR}^\mathrm{rms}}\hat{Q}_{2\parallel}^*\text{\ with }\hat{Q}_{2\parallel}^*=\frac{\sqrt{2}}{\pi}f_\mathrm{R}\left|\frac{\partial f_\mathrm{L}}{\partial V_\mathrm{LR}}\right|^{-1}\,.
\end{align}
\end{subequations}
$V_\mathrm{ac}$ is the amplitude of the drive and we have lumped electric field fluctuations into effective rms gate voltage fluctuations $\delta V_\mathrm{C}^\mathrm{rms}$ and $\delta V_\mathrm{LR}^\mathrm{rms}$. We assume here that the Rabi oscillations are driven with opposite modulations on the L and R gates. In the above expressions, both the Rabi frequency $f_\mathrm{R}$ and the LSESs $\partial f_\mathrm{L}/\partial V$ are given in MHz/mV/T (or equivalent unit). In the following, we focus the discussion on the normalized quality factors $\hat{Q}_{2\parallel,\perp}^*$, $\delta V_\mathrm{C}^\mathrm{rms}$ and $\delta V_\mathrm{LR}^\mathrm{rms}$ being strongly dependent on device layout and quality.
}

\begin{figure}
\centering
\includegraphics[width=.45\columnwidth]{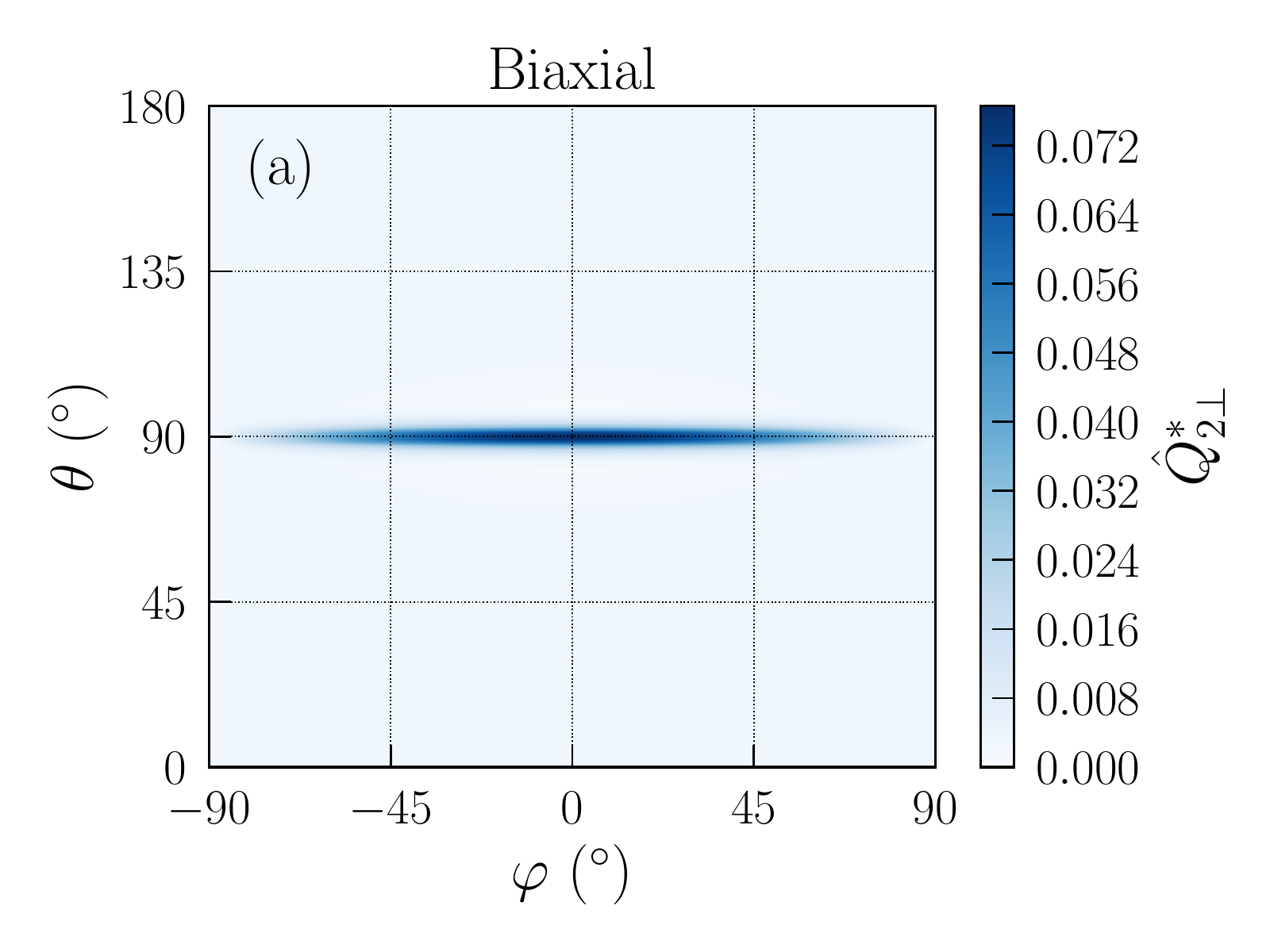}
\includegraphics[width=.45\columnwidth]{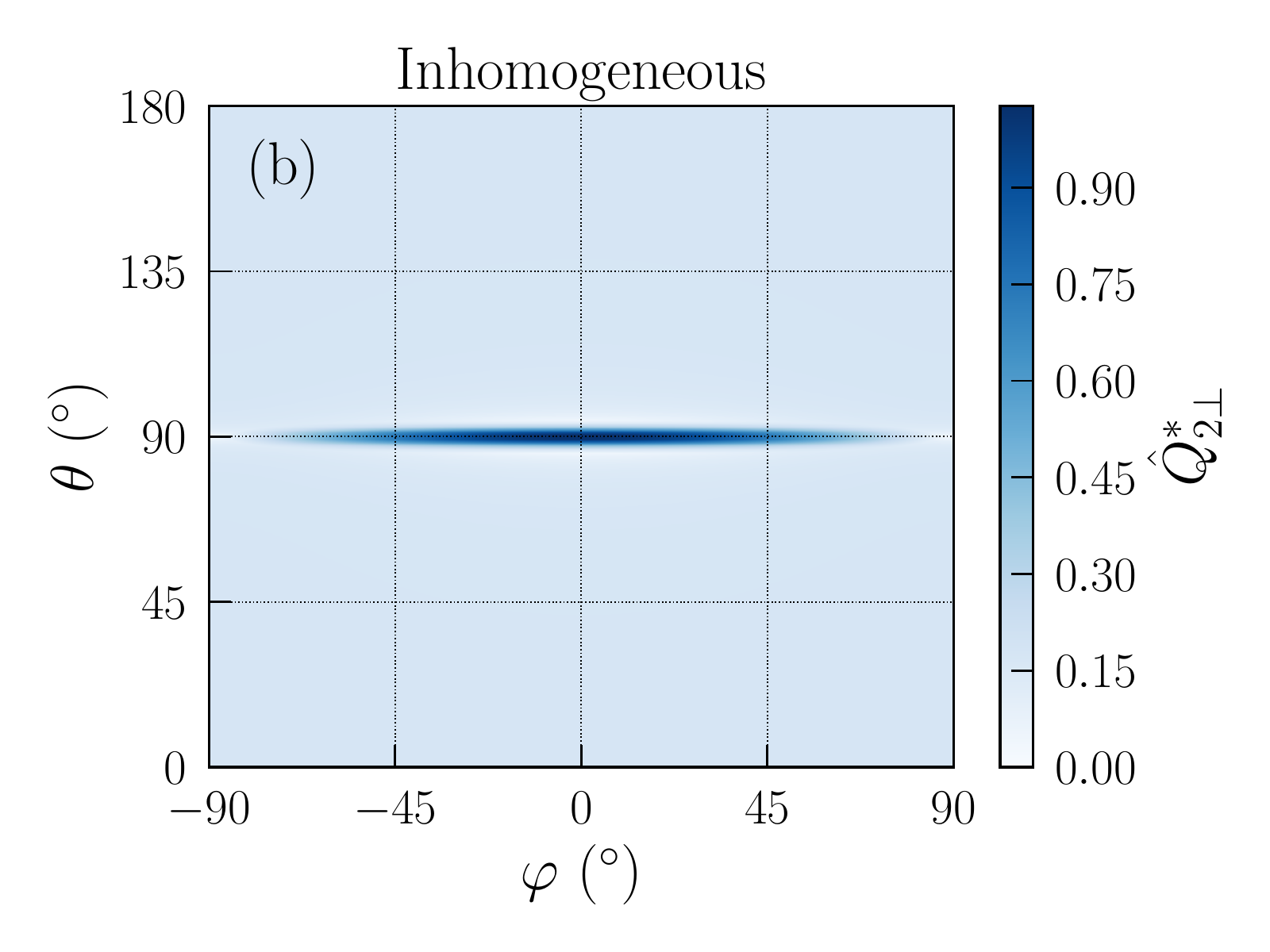}
\caption{\YMN{(a, b) Maps of the normalized quality factor $\hat{Q}_{2\perp}^*$ as a function of the orientation of the magnetic field, for opposite drives $\delta V_\mathrm{L}=-\delta V_\mathrm{R}=\tfrac{1}{2}V_\mathrm{ac}\cos(2\pi f_\mathrm{L}t)$ on the L and R gates ($V_\mathrm{C}=-40$\,mV). The top Ge$_{0.8}$Si$_{0.2}$ barrier is 50 nm thick. Map (a) is for homogeneous biaxial strains, and map (b) is with inhomogeneous TC strains.}}
\label{fig:Q2perp}
\end{figure}

\YMN{The map of $\hat{Q}_{2\perp}^*$ is plotted as a function of the orientation of the magnetic field in Fig.~\ref{fig:Q2perp}, for both biaxial and inhomogeneous strains ($V_\mathrm{C}=-40$\,mV). We emphasize that there are no sweet spots in the LSES $\partial f_\mathrm{L}/\partial V_\mathrm{C}$, neither at this bias nor at any other in the range $V_\mathrm{C}\in[-20,\,-200]$\,mV. Therefore, the hole never completely decouples from vertical electric field noise. Actually, the LSES increases monotonously from in-plane to vertical magnetic fields. As the Rabi frequency is maximal for $\mathbf{B}\parallel\mathbf{x}$, the quality factor $\hat{Q}_{2\perp}^*$ peaks there. Strinkingly, $\hat{Q}_{2\perp}^*$ is much larger in inhomogeneous than in biaxial strains. Indeed, the LSES $\partial f_\mathrm{L}/\partial V_\mathrm{C}$ is almost the same in both cases, while the Rabi frequency is enhanced by more than one order of magnitude in inhomogeneous strains. This is reminiscent of Fig.~\ref{fig:driveC}: for symmetry reasons, the strain-induced SOI is hardly harnessed by modulations of $V_\mathrm{C}$, neither in the transverse (Rabi) nor longitudinal (LSES) susceptibilities. As a consequence, the qubit is more resilient to vertical electric field fluctuations when inhomogeneous strains speed up electrical manipulation.}

\begin{figure}
\centering
\includegraphics[width=.45\columnwidth]{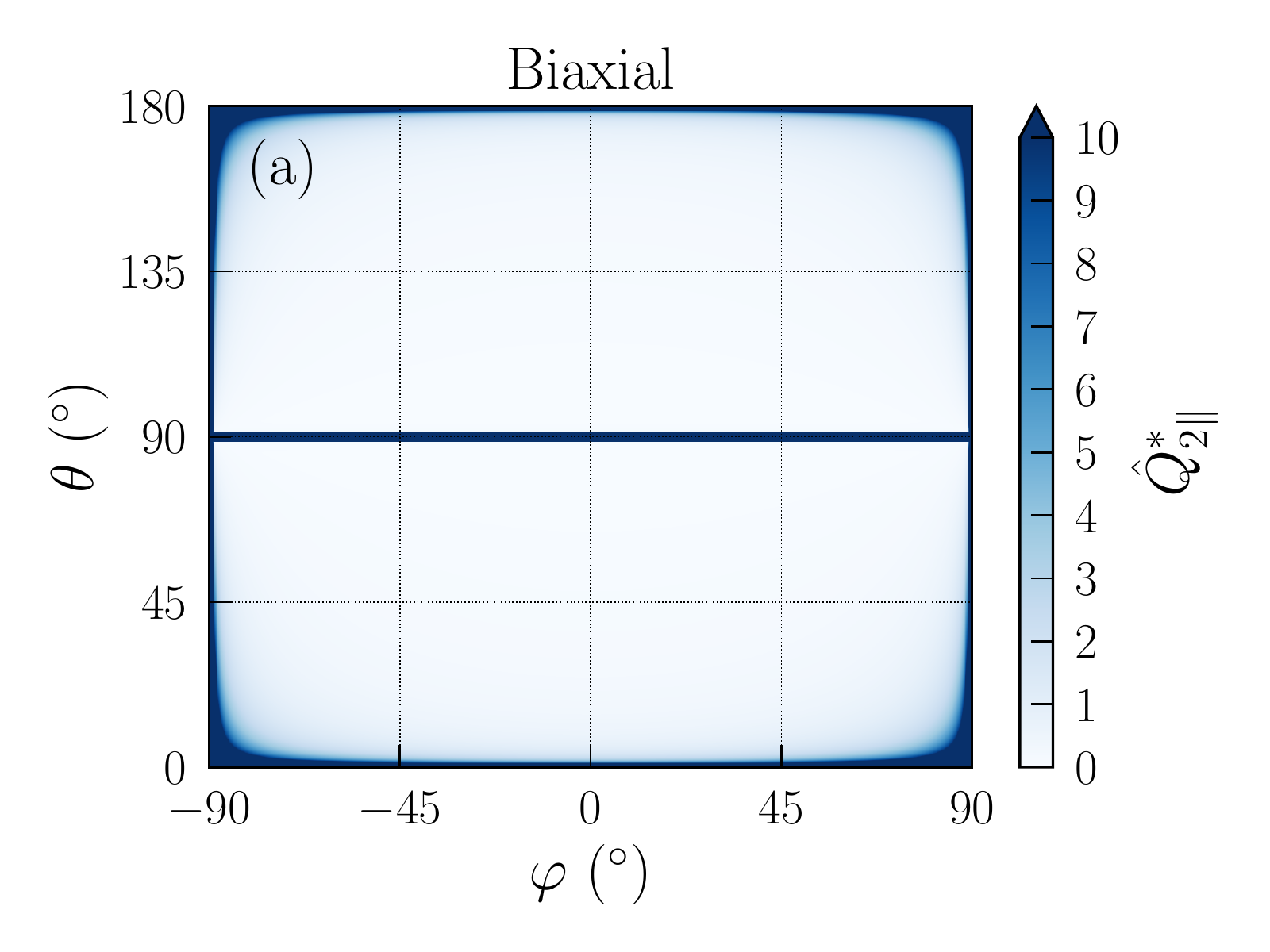}
\includegraphics[width=.45\columnwidth]{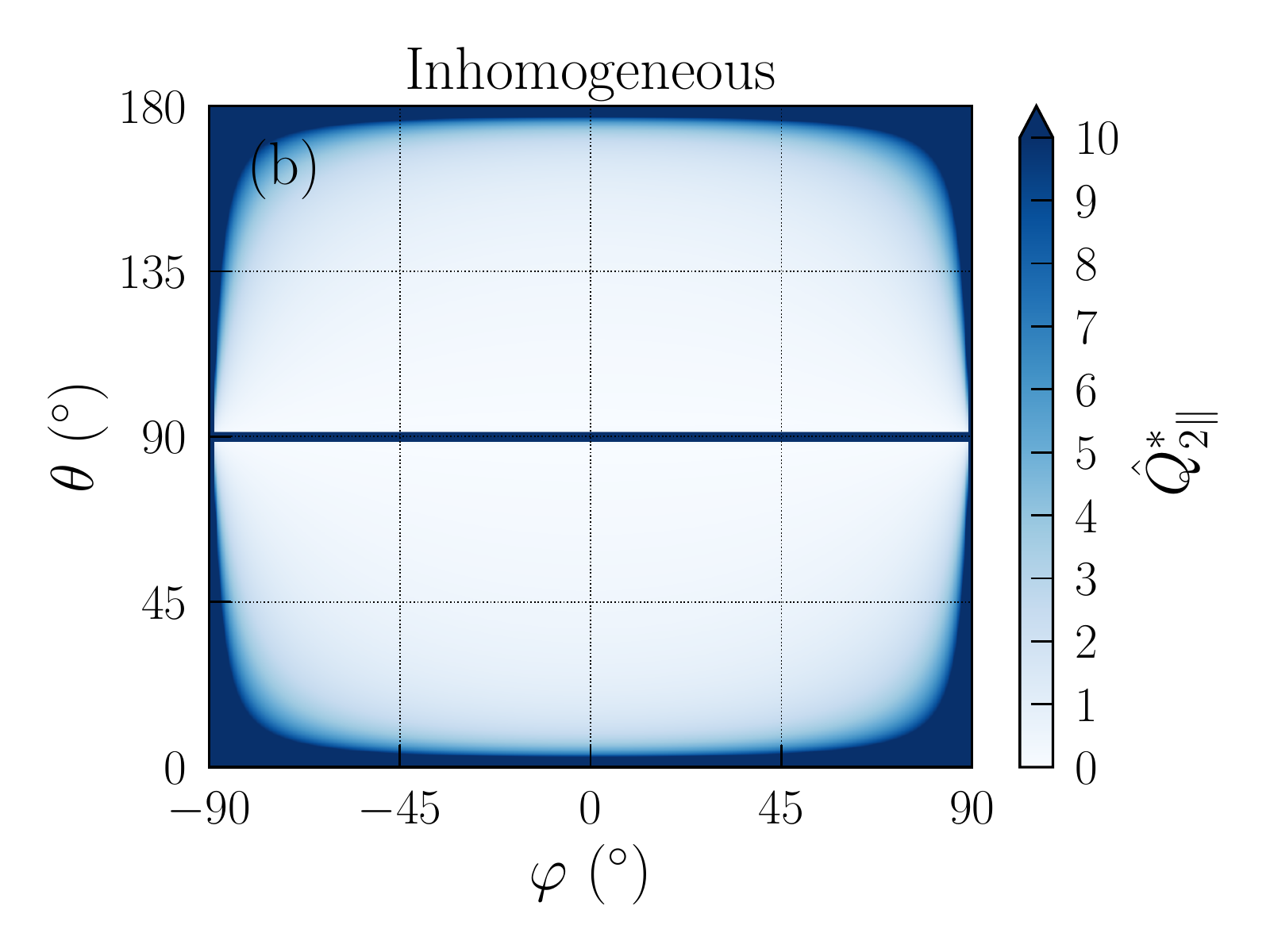}
\caption{\YMN{(a, b) Maps of the normalized quality factor $\hat{Q}_{2\parallel}^*$ as a function of the orientation of the magnetic field (same conditions as in Fig.~\ref{fig:Q2perp}). Map (a) is for homogeneous biaxial strains, and map (b) is with inhomogeneous TC strains.}}
\label{fig:Q2par}
\end{figure}

\YMN{The map of $\hat{Q}_{2\parallel}^*$ is likewise plotted in Fig.~\ref{fig:Q2par}. In that case, there is a clear sweet spot for $\mathbf{B}\parallel\mathbf{z}$ and a whole sweet line for in-plane magnetic fields. At this sweet spot and along this line, the qubit is decoupled (to first-order) from electrical noise. As discussed above, the Rabi frequency maxima lie at the sweet spot ($\mathbf{B}\parallel\mathbf{z}$) and along the sweet line ($\mathbf{B}\parallel\mathbf{x}$) due to the reciprocal sweetness between the transverse and longitudinal spin susceptibilities of the same gates \cite{michal2022tunable}. Again, the quality factors are (slightly) better in inhomogeneous strains, which highlights that stronger SOI does not necessarily degrade the figures of merit of the qubit.}

\YMN{The fastest manipulation and the best quality factors $\hat{Q}_{2\parallel}^*$ and $\hat{Q}_{2\perp}^*$ are, therefore, both achieved when setting $\mathbf{B}\parallel\mathbf{x}$, and in inhomogeneous strains. \YMNN{We have also computed the relaxation rates due to single-phonon emission at $T=100$\,mK and $f_\mathrm{L}=5$\,GHz along the lines of Ref.~\cite{Li20}. The relaxation times for $\mathbf{B}\parallel\mathbf{x}$ are almost the same in inhomogeneous strains ($T_1=535$\,$\mu$s) as in biaxial strains ($T_1=479$\,$\mu$s), despite the enhancement of spin-orbit coupling. The coupling of hole spins to phonons through uniaxial and shear deformation potentials indeed follows different trends than the coupling to electric fields \cite{Li20}}. We emphasize, though, that the magnetic field must be well aligned to make the most of these qubits. The sharpness of the in-plane features results from the strong anisotropy between $\gt_\parallel$ and $\gt_\perp$. We defer to a later publication an in-depth discussion about the engineering of the $\gt$-factor anisotropy and about the decoupling to vertical electric field noise, which are both non specific to strain-induced spin-orbit interactions.}

\section{Numerical $\gt$-matrices and gauge invariance}
\label{sec:gauge}

We first discuss the calculation of the numerical $\gt$-matrices used to produce Fig.~2 of the main text as well as Figs.~\ref{fig:driveL}-\ref{fig:Q2par}, then gauge invariance in the $\gt$-matrix formalism.

Let $H(\mathbf{B})$ be the Hamiltonian of the system (for an arbitrary choice of gauge) and let $\{\ket{\Uparrow},\,\ket{\Downarrow}\}$ be the ground-state doublet at $\mathbf{B}=\mathbf{0}$, with energy $E_0$. The $\gt$-matrix in the $\{\ket{\Uparrow},\,\ket{\Downarrow}\}$ basis set can be written \cite{Venitucci18}:
\begin{equation}
\gt=-\frac{2}{\mu_B}
\begin{pmatrix}{}
{\rm Re}\bra{\Downarrow}M^{x}\ket{\Uparrow} & {\rm Re}\bra{\Downarrow}M^{y}\ket{\Uparrow} & {\rm Re}\bra{\Downarrow}M^{z}\ket{\Uparrow} \\
{\rm Im}\bra{\Downarrow}M^{x}\ket{\Uparrow} & {\rm Im}\bra{\Downarrow}M^{y}\ket{\Uparrow} & {\rm Im}\bra{\Downarrow}M^{z}\ket{\Uparrow} \\
\bra{\Uparrow}M^{x}\ket{\Uparrow} & \bra{\Uparrow}M^{y}\ket{\Uparrow} & \bra{\Uparrow}M^{z}\ket{\Uparrow} \\
\end{pmatrix}\,,
\label{eq:gmatrix}
\end{equation}
where
\begin{equation}
M^\alpha=-\left.\frac{\partial H}{\partial B_\alpha}\right|_{\mathbf{B}=\mathbf{0}}   
\end{equation}
is the derivative of the Hamiltonian with respect to the magnetic field along $\alpha\in\{x\,,y\,,z\}$. Practically, the numerical $\gt$-matrices are computed from Eq.~(\ref{eq:gmatrix}) with the finite-difference ground-state wave functions of the LK Hamiltonian at $\mathbf{B}=\mathbf{0}$ \cite{Venitucci18}. These numerical $\gt$-matrices are, therefore, non-perturbative, at variance with the $\delta\gt_{\alpha\beta}$'s and SOI interactions obtained from the Schrieffer-Wolff transformation [Eqs.~(7) and (\ref{eq:calphabeta})].

We emphasize, though, that the choice of basis set $\{\ket{\Uparrow}\,,\ket{\Downarrow}\}$ is not unique as these states are degenerate. We remind that a rotation of the basis set does change the $\gt$-matrix but not the observables such as the Larmor and Rabi frequencies. Indeed, the rotated basis set $\{\ket{\tilde\Uparrow}\,,\ket{\tilde\Downarrow}\}$ can be related to the original basis set $\{\ket{\Uparrow}\,,\ket{\Downarrow}\}$ by a unitary $2\times2$ matrix $R$:
\begin{equation}
\begin{pmatrix}{}
\ket{\tilde\Uparrow} \\ 
\ket{\tilde\Downarrow}
\end{pmatrix}=R
\begin{pmatrix}{}
\ket{\Uparrow} \\ 
\ket{\Downarrow}
\end{pmatrix}\,.
\end{equation}
As discussed in Ref.~\onlinecite{Venitucci18}, $R$ can be further associated with a real, unitary $3\times 3$ matrix $U(R)$ such that the $\gt$-matrix $\tilde\gt(R)$ reads in the new basis set:
\begin{equation}
\tilde\gt(R)=U(R)\gt\,.
\end{equation}  
This transformation preserves the effective $\gt$-factor $\gt^*$ and the Larmor frequency $f_\mathrm{L}=\gt^*\mu_BB/h$:
\begin{equation}
\gt^*=|(U\gt)\mathbf{b}|=|\gt\mathbf{b}|    
\end{equation}
as well as the Rabi frequency:
\begin{equation}
f_\mathrm{R}=\frac{\mu_BBV_\mathrm{ac}}{2h\gt^*}\left|(U\gt)\mathbf{b}\times (U\gt^\prime)\mathbf{b}\right|=\frac{\mu_BBV_\mathrm{ac}}{2h\gt^*}\left|\mathrm{det}(U)U(\gt\mathbf{b}\times\gt^\prime\mathbf{b})\right|=\frac{\mu_BBV_\mathrm{ac}}{2h\gt^*}\left |\gt\mathbf{b}\times\gt^\prime\mathbf{b}\right|\,.
\end{equation} 

The derivative $\gt^\prime(V_\mathrm{G}=V_0)$ with respect to a given gate voltage $V_\mathrm{G}$ is computed by finite differences between two bias points $V_\mathrm{G}=V_0+\delta V$ and $V_\mathrm{G}=V_0-\delta V$. Care must be taken in ensuring a consistent choice of basis sets at the three bias points $V_\mathrm{G}=V_0$, $V_\mathrm{G}=V_0+\delta V$ and $V_\mathrm{G}=V_0-\delta V$ \cite{Venitucci18}. The $\gt$-matrix $\gt(V_\mathrm{G}=V_0)$ is finally diagonalized by a singular value decomposition, and $\gt^\prime(V_\mathrm{G}=V_0)$ is transformed accordingly \cite{Venitucci18}. As discussed above, this change of basis set has no impact on the Larmor and Rabi frequencies; however, the symmetry patterns of Table~\ref{tab:symmetries} actually apply (and thus can only be verified) in the basis set where $\gt(V_\mathrm{G}=V_0)$ is diagonal. 

We now discuss gauge invariance in the $\gt$-matrix formalism. Under a change of gauge $\mathbf{A}\to\mathbf{A}_\Omega=\mathbf{A}-\boldsymbol{\nabla}\Omega(\mathbf{r})$, where $\Omega(\mathbf{r})$ may depend on $\mathbf{B}$, the Hamiltonian transforms as:
\begin{equation}
H\to H_\Omega=T_\Omega HT_\Omega^\dagger\,;\,\,T_\Omega=e^{ie\Omega(\mathbf{r})/\hbar}\,.
\end{equation}
$H_\Omega$ and $H$ describe the same physics and share, therefore, the same spectrum; the ground states of $H_\Omega(\mathbf{B}=\mathbf{0})$ with energy $E_0$ are simply $\ket{\Uparrow}_\Omega=T_\Omega^0\ket{\Uparrow}$ and $\ket{\Downarrow}_\Omega=T_\Omega^0\ket{\Downarrow}$, where $T_\Omega^0= T_\Omega(\mathbf{B}=\mathbf{0})$. We can next introduce the operator:
\begin{equation}
M_\Omega^\alpha=-\left.\frac{\partial H_\Omega}{\partial B_\alpha}\right|_{\mathbf{B}=\mathbf{0}}=T_\Omega^0 M^\alpha T_\Omega^{0\dagger}-i\frac{e}{\hbar}\left[\left.\frac{\partial\Omega(\mathbf{r})}{\partial B_\alpha}\right|_{\mathbf{B}=\mathbf{0}},H_\Omega(\mathbf{B}=\mathbf{0})\right]\,,
\label{eq:Momega}
\end{equation}
and compute the $\gt$-matrix $\gt_\Omega$ in the $\{\ket{\Uparrow}_\Omega\,,\ket{\Downarrow}_\Omega\}$ basis set from Eq.~(\ref{eq:gmatrix}). As $\ket{\Uparrow}_\Omega$ and $\ket{\Downarrow}_\Omega$ are both eigenstates of $H_\Omega(\mathbf{B}=\mathbf{0})$ for the same energy $E_0$, the above commutator does not contribute, and we reach immediately:
\begin{equation}
\gt_\Omega=\gt\,.
\end{equation}
Therefore, the $\gt$-matrix is the same in the new gauge (in the corresponding gauge-dependent basis set). Whenever $\Omega(\mathbf{r})\propto B_x,\,B_y,\,B_z$ (which is the case when switching, e.g., between a symmetric and a Landau-type gauge), $T_\Omega^0=\mathbb{1}$ (the Hamiltonians are the same at $\mathbf{B}=\mathbf{0}$) and $\ket{\Uparrow}_\Omega\equiv\ket{\Uparrow}$, $\ket{\Downarrow}_\Omega\equiv\ket{\Downarrow}$: the $\gt$-matrix can be computed in the same basis set in the original and new gauges, and is invariant. Given the role of the magnetic vector potential in the Rashba interactions, we have carefully checked that the numerical $\gt$-matrices and Rabi frequencies are indeed gauge-invariant (within $0.1\%$ owing to the finite-difference discretization). We have also compared the Rabi frequencies computed in the $\gt$-matrix formalism with direct evaluations of the electric dipole matrix elements at finite magnetic field \cite{Venitucci18} in order to validate the computational results.

The equations (\ref{eq:calphabeta}) that result from a perturbation theory are gauge-invariant because they only involve the generalized momentum $\mathbf{p}=-i\hbar\boldsymbol{\nabla}+e\mathbf{A}$ that transforms according to $T_\Omega\mathbf{p}T_\Omega^\dagger=-i\hbar\boldsymbol{\nabla}+e\mathbf{A}_\Omega$.

\section{Effect of the Rashba interaction in the $\gt$-matrix formalism}
\label{sec:rashba}

We discuss the effect of a Rashba interaction $\delta{\cal H}_\mathrm{so}=-\hbar( p_x\sigma_y-p_y\sigma_x)/(m_\parallel\ell_\mathrm{so})$ on the $\gt$-matrix of a hole driven along $x$. In the absence of HH-LH mixing, the Hamiltonian of the heavy-hole envelopes reads at $\mathbf{B}=\mathbf{0}$:
\begin{equation}
{\cal H}_\mathrm{HH}=P_\mathrm{K}+Q_\mathrm{K}+V_\mathrm{t}\,,
\end{equation}
where $V_\mathrm{t}=V-a_v(\varepsilon_{xx}+\varepsilon_{yy}+\varepsilon_{zz})-\frac{1}{2}b_v(\varepsilon_{xx}+\varepsilon_{yy}-2\varepsilon_{zz})$ is the total potential. We assume for the sake of demonstration that $V_\mathrm{t}(\mathbf{r})\equiv V_x(x)+V_y(y)+V_z(z)$ is separable in the $x$, $y$, $z$ coordinates and that $V_x(x)\approx\tfrac{1}{2}Kx^2$ is roughly harmonic within the dot. Dealing with the HH-LH couplings to first-order in $\Delta_\mathrm{LH}$, the effective Hamiltonian for motion along $x$ is
\begin{equation}
{\cal H}_x=\frac{p_x^2}{2m_\parallel}+\frac{1}{2}Kx^2-\frac{\hbar}{m_\parallel\ell_\mathrm{so}}p_x\sigma_y+\frac{1}{2}\mu_B\boldsymbol{\sigma}\cdot\gt\mathbf{B}
\label{eq:linrashba}
\end{equation}
with $m_\parallel=m_0/(\gamma_1+\gamma_2-\gamma_h)$ the in-plane HH mass \cite{Michal21} and $p_x=-i\hbar\tfrac{\partial}{\partial x}+eA_x$. Here $\gt$ is the $\gt$-matrix of the ground-state doublet we are interested in, whose elements are given by Eqs.~(6) and (7) of the main text (it is not necessary to account for different $\gt$-matrices for excited Kramers pairs at lowest order).

In order to calculate $\gt^\prime=\partial\gt/\partial V_\mathrm{G}$, we assume that the gate G creates a homogeneous electric field ${\cal E}_\mathrm{d}\propto\delta V_\mathrm{G}$ along $x$, and thus add a (for now static) driving term $V_\mathrm{d}(\mathbf{r})=-e{\cal E}_\mathrm{d}x$ to the above Hamiltonian. Such a homogeneous electric field simply translates the dot as a whole by $x_\mathrm{d}(\delta V_\mathrm{G})=e{\cal E}_\mathrm{d}/K$. We need, however, to deal carefully with the effects of the Rashba interaction along this translation. For that purpose, we eliminate the $\propto p_x\sigma_y$ term from the Hamiltonian with a unitary transformation $\tilde{\cal H}_x=e^WH_xe^{-W}$, where \cite{Michal21,Golovach06}:
\begin{equation}
W=-i\frac{x}{\ell_\mathrm{so}}\sigma_y-\frac{\mu_B}{ K\ell_\mathrm{so}}\epsilon_{\alpha\beta y}(\gt\mathbf{B})_\alpha\sigma_\beta\frac{\partial}{\partial x}\,.
\label{eq:S}
\end{equation}
Here $\epsilon$ is the Levi-Civita antisymmetric tensor and the sum over $\alpha,\beta\in\{x,\,y,\,z\}$ is implied. This yields, to first-order in $\mathbf{B}$ and $\ell_\mathrm{so}^{-1}$:
\begin{equation}
\tilde{\cal H}_x\equiv{\cal H}_x+\left[W,\,{\cal H}_x\right]= -\frac{\hbar^2}{2m_\parallel}\frac{\partial^2}{\partial x^2}-\frac{ie\hbar}{m_\parallel}A_x\frac{\partial}{\partial x}+\frac{1}{2}Kx^2-e{\cal E}_\mathrm{d}x+\frac{1}{2}\mu_B\boldsymbol{\sigma}\cdot\gt\mathbf{B}+\delta\tilde{\cal H}_\mathrm{Z}({\cal E}_\mathrm{d})\,, 
\label{eq:Htildex}
\end{equation}
where:
\begin{align}
\delta\tilde{\cal H}_\mathrm{Z}({\cal E}_\mathrm{d})&=\left[W,\,-e{\cal E}_\mathrm{d}x\right] \nonumber \\
&=\frac{e{\cal E}_\mathrm{d}}{K\ell_\mathrm{so}}\mu_B\epsilon_{\alpha\beta y }(\gt\mathbf{B})_\alpha\sigma_\beta \nonumber \\
&=\frac{x_\mathrm{d}}{\ell_\mathrm{so}}\mu_B\left[(\gt\mathbf{B})_z\sigma_x-(\gt\mathbf{B})_x\sigma_z\right] \nonumber \\
&=\frac{1}{2}\mu_B\left(2\gt_\perp\frac{x_\mathrm{d}}{\ell_\mathrm{so}}B_z\sigma_x-2\gt_\parallel\frac{x_\mathrm{d}}{\ell_\mathrm{so}}B_x\sigma_z\right)\,.
\label{eq:deltaHz}
\end{align}
The transformed Hamiltonian $\tilde{\cal H}_x$ does not couple spin to momentum any more (only to position). The operator $W$ may not commute with the Hamiltonian for motion along $y$ and $z$ because $\tfrac{\partial}{\partial x}$ does, in general, act on the vector potential components $A_y$ and $A_z$; however this gives rise to ${\cal O}(B^2)$ corrections that are irrelevant for the linear response $\gt$-matrices. The $\propto p_y\sigma_x$ term can be eliminated with a similar unitary transform but does not contribute to $\delta\tilde{\cal H}_\mathrm{Z}$ when the dot is driven along $x$. We can, therefore, compute the dressed $\gt$-matrix $\tilde\gt$ and its derivative from Eqs.~(\ref{eq:gmatrix}) and (\ref{eq:Htildex}) using the (separable) wave functions at $\mathbf{B}=\mathbf{0}$; from the above expressions we readily identify $\tilde\gt\equiv\gt$, $\gt_{zx}^\prime=-2\gt_\parallel x_\mathrm{d}^\prime/\ell_\mathrm{so}$ and $\gt_{xz}^\prime=2\gt_\perp x_\mathrm{d}^\prime/\ell_\mathrm{so}$ as stated in the main text. When the dot is driven resonantly with an AC signal $\delta V_\mathrm{G}=V_\mathrm{ac}\cos(2\pi f_\mathrm{L}t)$, we then reach:
\begin{equation}
f_\mathrm{R}=\frac{\mu_BBV_\mathrm{ac}}{2h\gt^*}\left |\gt\mathbf{b}\times\gt^\prime\mathbf{b}\right|=\frac{x_\mathrm{d}(V_\mathrm{ac})}{\ell_\mathrm{so}}f_\mathrm{L}
\end{equation}
when $\mathbf{B}$ lies in the $xz$ plane. This can be interpreted as the action of the time-dependent Rashba Hamiltonian:
\begin{equation}
{\cal H}_\mathrm{eff}(t)=-\frac{\hbar}{m_\parallel\ell_\mathrm{so}}p_x(t)\sigma_y=-\frac{\hbar}{\ell_\mathrm{so}}v_x(t)\sigma_y\equiv\left[hf_\mathrm{L}\frac{x_\mathrm{d}(V_\mathrm{ac})}{\ell_\mathrm{so}}\sin(2\pi f_\mathrm{L}t)\right]\sigma_y
\end{equation}
with $v_x(t)=-2\pi f_\mathrm{L}x_\mathrm{d}(V_\mathrm{ac})\sin(2\pi f_\mathrm{L}t)$ the classical velocity of the dot, or as the action of the effective time-dependent magnetic field:
\begin{equation}
{\cal B}_y(t)= \frac{2hf_\mathrm{L}}{\mu_B\gt_\perp}\frac{x_\mathrm{d}(V_\mathrm{ac})}{\ell_\mathrm{so}}\sin(2\pi f_\mathrm{L}t)\,.
\end{equation}

The unitary transformation Eq.~(\ref{eq:S}) only holds for a constant spin-orbit length $\ell_\mathrm{so}$. If $\ell_\mathrm{so}$ depends on position, the $\propto p_x\sigma_y$ interaction couples, in particular, the orbital motion of the hole in the magnetic vector potential to the position of the dot through the $-e\hbar A_x\sigma_y/[m_\parallel\ell_\mathrm{so}(\mathbf{r})]$ term (and so may the $\propto p_y\sigma_x$ interaction that was irrelevant for homogeneous motion along $x$). This coupling is actually canceled when $\ell_\mathrm{so}$ is a constant by the $[W, p_x^2/(2m_\parallel)]$ commutator in the unitary transform, Eq.~(\ref{eq:Htildex}). It gives rise to significant corrections to Eq.~(\ref{eq:deltaHz}) that can be evidenced by disabling the magnetic vector potential in the simulations, as highlighted in the main text. 

\bibliography{arxiv}

\end{document}